\documentclass[aps,prd,onecolumn,10pt,nofootinbib,floatfix]{revtex4-2}
\usepackage{amsmath,amssymb,mathtools,bm,booktabs,graphicx,microtype,mathrsfs}
\usepackage{xcolor}
\usepackage[T1]{fontenc}
\usepackage[utf8]{inputenc}
\usepackage{lmodern}
\usepackage{tikz}
\usepackage{float}
\usetikzlibrary{calc,arrows.meta,positioning}
\usepackage{comment}
\usepackage[colorlinks=true,citecolor=blue,linkcolor=blue,urlcolor=blue]{hyperref}
\allowdisplaybreaks
\hypersetup{colorlinks=true,citecolor=blue!60!black,urlcolor=blue!60!black,linkcolor=blue!60!black}
\newcommand{\D}{\mathrm D}

\newcommand{\Mom}{\mathcal P}
\newcommand{\CA}{\mathscr C}

\newcommand{\dd}{\mathrm d}
\allowdisplaybreaks[2]

\definecolor{lime}{HTML}{A6CE39}
\DeclareRobustCommand{\orcidicon}{%
 \begin{tikzpicture}
 \draw[lime, fill=lime] (0,0) 
 circle [radius=0.16] 
 node[white] {{\fontfamily{qag}\selectfont \tiny ID}};
 \draw[white, fill=white] (-0.0625,0.095) 
 circle [radius=0.007];
 \end{tikzpicture}
 \hspace{-2mm}
}

\foreach \x in {A, ..., Z}{%
 \expandafter\xdef\csname orcid\x\endcsname{\noexpand\href{https://orcid.org/\csname orcidauthor\x\endcsname}{\noexpand\orcidicon}}
}

\newcommand\orcidDavid{{\href{https://orcid.org/0009-0006-8485-3140}{\orcidicon}}}

\begin{document}

\title{Scalar-tensor interactions selected through their first-order thermodynamic dynamics}

\author{David S. Pereira\orcidDavid\!\!}
\email{djpereira@ciencias.ulisboa.pt}

\affiliation{Departamento de F\'{i}sica, Faculdade de Ci\^{e}ncias da Universidade de Lisboa, Campo Grande, Edif\'{i}cio C8, P-1749-016 Lisbon, Portugal}
\affiliation{Instituto de Astrof\'{i}sica e Ci\^{e}ncias do Espa\c{c}o, Faculdade de Ci\^{e}ncias da Universidade de Lisboa, Campo Grande, P-1749-016 Lisbon, Portugal}

\date{\today}

\begin{abstract}
We investigate the restrictive capability of first-order thermodynamics on scalar-tensor gravity. Starting from a general action containing nonlinear curvature dependence, an Einstein-tensor-Hessian coupling, and independent quadratic and cubic scalar-Hessian interactions, we derive the complete effective-fluid quantities. Simultaneously imposing Eckart's heat-conduction and shear-viscosity equations, while accounting for cancellations between interactions, substantially reduces the original operator basis and determines the geometric shear viscosity. Retaining the complete heat-conduction law, we establish an integrability condition that constrains the nonlinear Hessian interactions without prescribing the temperature and thermal conductivity. We construct explicit constitutive pairs and derive the associated thermal evolution, revealing contributions beyond the usual acceleration-aligned description. Remarkably, the thermodynamic integrability condition simplifies the scalar-metric kinetic structure such that imposing covariant degeneracy excludes the remaining nonlinear interactions, recovering the established luminal Horndeski subclass. Nonlinear thermodynamically compatible interactions nevertheless survive when degeneracy is restricted to the scalar-unitary gauge. These results establish a connection between thermodynamic integrability and dynamical degeneracy, demonstrating the potential of first-order thermodynamics as an independent and complementary criterion for selecting gravitational interactions.
\end{abstract}
\maketitle

\tableofcontents
\clearpage

\section{Introduction}\label{sec:intro}

Scalar-tensor theories constitute a broad class of extensions of general relativity, ranging from the original Brans-Dicke model to Horndeski gravity and its higher-order generalizations~\cite{Brans:1961sx,Horndeski:1974wa,Deffayet:2011gz,Gleyzes:2014qga,Langlois:2015cwa,BenAchour:2016fzp,Kobayashi:2019hrl}. Their theoretical development and applications span a wide range of problems in cosmology, astrophysics, and gravitational physics (see Refs.~\cite{Faraoni:2004pi,CANTATA:2021asi,Kobayashi:2019hrl,Quiros:2019ktw,Clifton:2011jh} for comprehensive reviews). Alongside these developments, the effective-fluid approach has provided a complementary perspective on scalar-tensor gravity by expressing its gravitational dynamics in terms of relativistic fluid variables~\cite{Faraoni:2018qdr,Nucamendi:2019uen,Faraoni:2022gry}. When the scalar field has a timelike gradient, it defines a preferred congruence of observers relative to which the gravitational field equations can be decomposed into an effective energy density, pressure, heat flux, and anisotropic stress~\cite{Faraoni:2018qdr,Nucamendi:2019uen,Faraoni:2021lfc,Faraoni:2022gry,Faraoni:2021jri,Giusti:2021sku}.

Moreover, beyond this effective-fluid description, its physical meaning can be further developed by establishing a correspondence with the constitutive relations of Eckart's first-order relativistic thermodynamics~\cite{Eckart:1940te}. In this formulation, the heat current is related to the spatial temperature gradient and fluid acceleration, while the trace-free viscous stress is proportional to the shear. Matching these relations with the effective-fluid quantities of scalar-tensor gravity provides a thermodynamic interpretation of the geometric heat flux and viscous stresses, together with the identification of effective gravitational temperature and transport coefficients. This correspondence has led to the development of the first-order thermodynamics of scalar-tensor gravity~\cite{Faraoni:2021lfc,Faraoni:2021jri,Giusti:2021sku,Giardino:2022sdv,Giardino:2023ygc}. The framework has since been investigated and extended in a variety of gravitational theories and physical settings, including cosmological configurations, Horndeski gravity, higher-curvature models, and the thermodynamic behavior of scalar gravitational degrees of freedom~\cite{Faraoni:2023hwu,Faraoni:2022doe,Miranda:2024dhw,Houle:2024sxs,Faraoni:2025fjq,Faraoni:2025alq,yhxk-dd5w,Miranda:2026ozq,Banerjee:2026ulx,Giusti:2026ymb,Pereira:2026hpi,Pereira:2026ccq,Pereira:2026xog,Pereira:2026cum}.

An important question, however, is whether the effective-fluid quantities of a general scalar-tensor theory can always be matched with Eckart's constitutive equations. Although the fluid decomposition can be constructed independently of such a matching, nonminimal curvature couplings and higher-derivative scalar interactions may generate heat-flow and anisotropic-stress contributions that are not compatible with the corresponding first-order relations. This issue has already been investigated in restricted classes of scalar-tensor gravity, where constitutive requirements have been shown to constrain Horndeski interactions~\cite{Miranda:2022wkz}, and kinetic dependence in nonminimal curvature couplings has been found to generate additional heat-flow components~\cite{Pereira:2025dmk}. Kinetic gravity braiding provides a related setting in which scalar-metric mixing produces a nontrivial effective energy current~\cite{Deffayet:2010qz,Pujolas:2011he}.

Motivated by these developments, we aim to establish the extent to which first-order thermodynamics can constrain scalar-tensor interactions beyond the gravitational theories for which such a correspondence has already been investigated. While previous studies have demonstrated that constitutive requirements can restrict Horndeski interactions~\cite{Miranda:2022wkz} and constrain the heat-flow contributions arising from particular curvature couplings~\cite{Pereira:2025dmk}, an important question remains as to whether these restrictions persist when a considerably broader set of gravitational interactions is allowed to contribute simultaneously. In particular, higher-derivative operators may compensate for contributions that would otherwise be incompatible with Eckart's constitutive equations. To address this question, we consider a broad class of scalar-tensor theories encompassing the complete set of independent interactions quadratic and cubic in second derivatives of the scalar field, together with nonlinear curvature dependence and an Einstein-tensor-Hessian coupling~\cite{Horndeski:1974wa,Deffayet:2011gz,Gleyzes:2014qga,Langlois:2015cwa,BenAchour:2016fzp,Langlois:2018dxi}. Neither the Horndeski structure nor the degeneracy conditions of higher-order scalar-tensor gravity are assumed at the outset. This allows us to assess the restrictive power of first-order thermodynamics independently of the conditions that ordinarily guide the construction of these theories.

To do such, we derive the complete effective energy density, pressure, heat flux, and anisotropic stress associated with the gravitational action. The heat flux and anisotropic stress are then compared simultaneously with Eckart's heat-conduction and shear-viscosity equations, retaining all possible cancellations between the different interaction sectors. Although the effective-fluid quantities are derived from the gravitational field equations, the constitutive matching is performed locally and off shell, without restricting the metric and scalar field to particular solutions or spacetime symmetries. As we shall see, the simultaneous application of these two constitutive relations is already sufficient to substantially reduce the original operator basis, even when compensating higher-derivative interactions are allowed. In particular, the analysis reveals that certain curvature contributions cannot be accommodated by Eckart's constitutive relations, independently of the freedom introduced by the quadratic and cubic scalar-Hessian operators.

Moreover, in the usual first-order thermodynamic description of scalar-tensor gravity, the geometric heat current is commonly directed along the acceleration of the scalar congruence. However, the higher-derivative interactions considered here allow for additional heat-flow directions (as it will be shown ahead), raising the question of whether such contributions can still admit a consistent first-order thermodynamic interpretation. Rather than excluding these contributions by requiring an acceleration-aligned heat current, we retain the complete Eckart heat-conduction equation and investigate the conditions under which a common effective temperature and thermal conductivity can account for the different spatial components of the geometric heat flux. We show that the existence of such a constitutive pair imposes an integrability condition directly on the Hessian dependence of the gravitational Lagrangian, leading to nontrivial relations between the independent quadratic and cubic coupling functions. This provides a further level of thermodynamic selection that cannot be obtained from the alignment of the heat current with the acceleration alone. Moreover, we examine the freedom in identifying the temperature and thermal conductivity, construct explicit constitutive pairs, and investigate the thermal evolution associated with the surviving nonlinear interactions. These interactions also allow us to explore a geometric thermal response that is no longer entirely governed by the nonminimal curvature coupling, in contrast with the corresponding shear viscosity.

By having established the restrictions arising from first-order thermodynamics, we complement them with the degeneracy conditions of higher-order scalar-tensor gravity~\cite{Langlois:2015cwa,BenAchour:2016fzp,Langlois:2018dxi}. The interest in combining these two independent requirements lies in the fact that the established DHOST classifications admit nonlinear scalar-Hessian interactions through cancellations between different operators in the coupled scalar-metric kinetic matrix. The restrictions imposed by Eckart's constitutive equations, however, substantially reduce the interactions available for such cancellations. We show that the thermodynamic integrability condition simplifies the kinetic structure of the selected theories to such an extent that requiring covariant degeneracy excludes their remaining nonlinear Hessian interactions. Interestingly, the combined application of these two criteria recovers the known subclass of Horndeski gravity with luminal gravitational-wave propagation, without imposing either the Horndeski structure or the gravitational-wave speed at the outset. This result is particularly relevant in view of the stringent observational constraints obtained from GW170817 and its electromagnetic counterpart~\cite{LIGOScientific:2017zic,Langlois:2017dyl}, together with recent investigations relating the geometric shear response to gravitational-wave propagation~\cite{Pereira:2026xog,Pereira:2026cum}. We also distinguish full covariant degeneracy from degeneracy restricted to the scalar-unitary gauge, as the latter admits a different class of nonlinear thermodynamically compatible interactions.

The purpose of this work is therefore not simply to extend the effective-fluid description to a broader class of scalar-tensor theories, nor to recover restrictions that have already been established within Horndeski gravity. Rather, we investigate how the complete first-order constitutive equations constrain a gravitational action in which the relations between the curvature and higher-derivative interactions have not been prescribed. In doing so, we establish a connection between the integrability of the geometric heat current, the functional structure of the gravitational Lagrangian, and the degeneracy of the coupled scalar-metric kinetic matrix. Although thermodynamic compatibility does not by itself establish dynamical consistency, perturbative stability, or observational viability, our results demonstrate that it provides a complementary criterion for selecting gravitational interactions before these additional physical requirements are imposed.

With this being said, this paper is organized as follows. Section~\ref{sec:DHOSTbasis} introduces the general scalar-tensor action and the effective-fluid frame. In Secs.~\ref{sec:metricvariation}--\ref{sec:fluidall}, we derive the gravitational field equations, construct the complete effective-fluid quantities, and introduce the constitutive relations used in the thermodynamic matching. Section~\ref{sec:matching} establishes the restrictions imposed on the gravitational action by the simultaneous matching of heat conduction and shear viscosity, while Sec.~\ref{sec:fullEckart} examines the integrability of the geometric heat current and the identification of the effective temperature and thermal conductivity. The corresponding thermodynamic sign conditions are discussed in Sec.~\ref{sec:thermalpositivity}. In Sec.~\ref{sec:health}, we examine the intersection between the thermodynamically selected interactions and the covariantly degenerate scalar-tensor theories, while Sec.~\ref{sec:thermalEvolution} investigates the thermal evolution associated with the nonlinear Hessian interactions and the corresponding evolution of the geometric shear viscosity. Section~\ref{sec:scope} discusses the scope of our results and their relation to equivalent representations of the gravitational action. Finally, in Sec.~\ref{sec:conclusions} we present our conclusions, with the technical details of the metric variation, fluid projections, and kinetic degeneracy collected in the appendices.

\section{The action and the scalar-tensor interactions }\label{sec:DHOSTbasis}
We work with metric signature $(-,+,+,+)$, $\phi_a=\nabla_a\phi$, $\phi_{ab}=\nabla_a\nabla_b\phi$ and $X=-\phi_a\phi^a/2>0$. The general action considered here is
\begin{equation}\label{eq:resummedaction}
S=\int\!\dd^4x\sqrt{-g}\left[f(\phi,X,R)+Q(\phi,X)\Box\phi+\sum_{i=1}^{5}A_i(\phi,X)L_i^{(2)}+F_{(3)}(\phi,X)G^{ab}\phi_{ab}+\sum_{j=1}^{10}B_j(\phi,X)L_j^{(3)}\right]+S_m[g,\psi_m].
\end{equation}
This action encompasses several of the most extensively investigated classes of scalar-tensor gravity, including Brans-Dicke theory, kinetic gravity braiding, covariant Galileons, the complete Horndeski family, and quadratic and cubic beyond-Horndeski and DHOST theories~\cite{Brans:1961sx,Horndeski:1974wa,Deffayet:2010qz,Deffayet:2011gz,Gleyzes:2014qga,Langlois:2015cwa,BenAchour:2016cay,BenAchour:2016fzp}. The coupling functions $A_i(\phi,X)$ and $B_j(\phi,X)$ correspond to the standard DHOST operator basis, comprising five quadratic and ten cubic scalar-Hessian invariants. However, arbitrary choices of these functions do not necessarily define a degenerate theory. We therefore keep all coupling functions independent, without imposing the relations that characterize Horndeski or DHOST theories. Furthermore, the function $f(\phi,X,R)$ extends the usual linear-curvature basis by allowing nonlinear dependence on the Ricci scalar, which may introduce an additional gravitational scalar degree of freedom~\cite{Sotiriou:2008rp}. This choice allows us to investigate the restrictions imposed by first-order thermodynamics on a considerably broader class of gravitational interactions, independently of the conditions that ordinarily determine their dynamical structure.

We restrict our analysis to the operator basis in Eq.~\eqref{eq:resummedaction}, without including curvature invariants other than functions of the Ricci scalar and the specified Einstein-tensor-Hessian coupling, derivatives of curvature, or interactions beyond cubic order in the scalar Hessian. Throughout this work, we adopt a fixed Jordan matter frame. When $f_{RR}=0$, we write the general function as $f=P(\phi,X)+F_{(2)}(\phi,X)R$ and refer to the resulting theory as the linear-curvature action. The corresponding redefinitions of the coupling functions under integration by parts are discussed in Sec.~\ref{sec:scope}. When comparing our results with the conventional DHOST formulation, we use $X_{\rm D}=\phi_a\phi^a=-2X$, so that $\partial_{X_{\rm D}}=-\tfrac12\partial_X$. The complete index contractions defining the quadratic and cubic operators are listed in Appendix~\ref{app:basisops}. In what follows, we express these operators through a reduced set of scalar-Hessian contractions, allowing their individual and combined contributions to the effective-fluid quantities to be identified more directly.

To avoid carrying the long indexed expressions through every variation, we define $x_a\equiv\phi_a$, $Y_{ab}\equiv\phi_{ab}$ and the six scalar contractions as
\begin{align}
 t&=Y^a{}_a=\Box\phi,& U&=Y_{ab}Y^{ab},& V&=Y^a{}_bY^b{}_cY^c{}_a,\notag\\
 S&=x^aY_{ab}x^b,& W&=x^aY_{ab}Y^{bc}x_c,& Z_3&=x^aY_{ab}Y^{bc}Y_{cd}x^d.\notag
\end{align}
With these expressions, the five quadratic and ten cubic invariants are given explicitly by (see Appendix~\ref{app:basisops}):
\begin{align}
 (L_1^{(2)},\ldots,L_5^{(2)})&=(U,t^2,tS,W,S^2),\notag\\
 (L_1^{(3)},\ldots,L_{10}^{(3)})&=(t^3,tU,V,t^2S,tW,US,\notag\\
 &\quad Z_3,SW,tS^2,S^3).
 \label{eq:basis}
\end{align}
For the algebraic differentiation below, we also set $v_a=Y_{ab}x^b$, $w_a=Y_{ab}v^b$ and $\ell_a=Y_{ab}w^b$. The scalar invariant $S=x^aY_{ab}x^b$ should not be confused with the action functional $S[g,\phi,\psi_m]$.

The quartic Horndeski Hessian identity takes the following form in our {positive}-$X$ convention as
\begin{equation}\label{eq:quarticdictionary}
 A_1=-F_{(2),X},\quad A_2=F_{(2),X},\quad A_3=A_4=A_5=0
 \ \Longrightarrow\quad
 \sum_{i=1}^5 A_iL_i^{(2)}=F_{(2),X}\big[(\Box\phi)^2-\phi_{ab}\phi^{ab}\big].
\end{equation}
In the alternative DHOST convention $X_{\rm D}=\phi_a\phi^a$, the same coefficients read $A_1=2F_{(2),X_{\rm D}}$ and $A_2=-2F_{(2),X_{\rm D}}$ \cite{Langlois:2018dxi,Lazanu:2024mzj}. Likewise $L_1^{(3)}-3L_2^{(3)}+2L_3^{(3)}$ is the familiar cubic Hessian combination entering the Horndeski $G_5$ sector, with convention-dependent derivative prefactors. These are comparison identities, not restrictions imposed on the general action. The $F_{(3)}G^{ab}\phi_{ab}$ curvature-Hessian coupling belongs to the standard cubic classification; for special $F_{(3)}$ its contribution can be rearranged by integration by parts, but it cannot be discarded for arbitrary $F_{(3)}(\phi,X)$. The non-invariance of naive boundary-term thermodynamic assignments is also discussed in Ref.~\cite{Bhattacharya:2024vbx}.

\subsection{1+3 decomposition}
For a timelike scalar-field gradient, we consider the standard covariant $1+3$ decomposition adapted to the scalar congruence~\cite{Faraoni:2021lfc}. Defining $s\equiv\sqrt{2X}$, the four-velocity, spatial projection tensor, and spatial derivative are given by
\begin{equation}
u_a\equiv\frac{x_a}{s},\quad
h_{ab}\equiv g_{ab}+u_au_b,\quad
\D_a\equiv h_a{}^b\nabla_b.
\end{equation}
Here, $u_a$ is the unit timelike four-velocity associated with the scalar gradient, $h_{ab}$ projects onto the three-dimensional hypersurfaces orthogonal to $u^a$, and $\D_a$ denotes the corresponding spatially projected covariant derivative. Furthermore, the acceleration, expansion tensor, expansion scalar and shear are given by
\begin{align}
 a_a&=u^b\nabla_bu_a,\quad K_{ab}=h_a{}^ch_b{}^d\nabla_cu_d,\notag\\
 \theta&=h^{ab}K_{ab},\quad \sigma_{ab}=K_{ab}-\tfrac13\theta h_{ab}.
\end{align}
Since $u_a$ is hypersurface orthogonal, its vorticity vanishes.

Differentiating $x_a=su_a$ and using the definitions of the scalar congruence, we introduce the quantities
\begin{equation}
d\equiv\dot s=u^a\nabla_a s,
\quad
t\equiv g^{ab}Y_{ab}=\Box\phi,
\end{equation}
where $d$ describes the evolution of the scalar-gradient magnitude along the fluid worldlines, while $t$ denotes the trace of the scalar Hessian. The corresponding kinematic identities are
\begin{align}
\dot\phi &=-s, & \D_a\phi &=0, \notag\\
\dot s &=d, & \D_as &=-sa_a, \notag\\
\dot X &=sd, & \D_aX &=-2Xa_a,
\end{align}
and, by using the symmetry $Y_{ab}=Y_{ba}$, the scalar Hessian can be decomposed as
\begin{equation}
\label{eq:hesskin}
Y_{ab}=sK_{ab}-d\,u_au_b-2s\,u_{(a}a_{b)},
\end{equation}
whose trace gives
\begin{equation}
t\equiv Y^a_a=d+s\theta,
\end{equation}
with $\theta=\nabla_a u^a$ denoting the expansion of the scalar congruence.

\section{Metric field equations and the effective stress tensor}\label{sec:metricvariation}

\subsection{Metric variation and the common source tensor}
A consistent comparison requires the complete variation of the action. In particular, changing the metric changes not only explicit index contractions but also the connection inside $\nabla_a\nabla_b\phi$. The latter part is where third derivatives enter the effective fluid, and it is also where cancellations between apparently different operators occur. For simplicity we write
\begin{equation}\label{eq:actionvariation}
\mathcal L_{\rm grav}=f(\phi,X,R)+\mathcal J_0+F_{(3)}G^{ab}Y_{ab},\quad
\mathcal J_0=Qt+\sum_{i=1}^{5}A_iL_i^{(2)}+\sum_{i=1}^{10}B_iL_i^{(3)}.
\end{equation}
The potential and general k-essence term are already included in $f$; on the linear-curvature branch $f=P+F_{(2)}R$. The scalar and metric are otherwise arbitrary.

At fixed $\phi$, variation of the inverse metric gives
\begin{align}\label{eq:fundamentalvariations}
\delta X&=-\tfrac12x_ax_b\delta g^{ab},&
\delta Y_{ab}&=-x_c\delta\Gamma^c{}_{ab},&
\delta R_{ab}&=\nabla_c\delta\Gamma^c{}_{ab}-\nabla_b\delta\Gamma^c{}_{ac}.
\end{align}
The $f$ variation includes $\delta R$ and derivatives of $f_R$, while each Hessian monomial contributes in two ways: an algebraic inverse-metric derivative and a derivative of the connection. The $F_{(3)}G^{ab}Y_{ab}$ interaction has, in addition, its own Ricci variation. For this purpose define the following fully specified response tensors:
\begin{align}
M_{ab}=\left.\frac{\partial(\mathcal J_0+F_{(3)}G^{cd}Y_{cd})}{\partial g^{ab}}\right|_{x,Y,R_{cd}}, \ 
\Mom^{ab}=\frac{\partial(\mathcal J_0+F_{(3)}G^{cd}Y_{cd})}{\partial Y_{ab}}, \
E^{ab}=F_{(3)}\left(Y^{ab}-\frac12t g^{ab}\right).\notag
\end{align}
These are algebraic derivatives of a given polynomial, not assumed effective-fluid functions. Their complete expressions are given in Eqs.~\eqref{eq:fullM}-\eqref{eq:momentumall} below. Substituting Eq.~\eqref{eq:fundamentalvariations}, integrating by parts and dropping the boundary variations yields
\begin{subequations}\label{eq:connectionvariations}
\begin{align}
I^Y_{ab}&=-\nabla_c\big(x_{(a}\Mom_{b)}{}^c\big)
+\frac12\nabla^c\big(x_c\Mom_{ab}\big),\label{eq:IY}\\
I^R_{ab}&=\frac12\Box E_{ab}-\nabla_c\nabla_{(a}E_{b)}{}^c
+\frac12g_{ab}\nabla_c\nabla_dE^{cd}.\label{eq:IR}
\end{align}
\end{subequations}
The order of the derivatives in the second term of $I^R_{ab}$ is important. It follows directly from integrating the Palatini variation twice; commuting the derivatives adds an explicit Riemann-Ricci contraction. Appendix~\ref{app:F3audit} derives all three contributions to the $F_{(3)}$ metric response without setting the background curvature or the derivatives of $F_{(3)}$ to zero.

There are two reasons to keep these expressions intact. First, even $Q\Box\phi$ has a connection variation: when $Q$ is constant it cancels the other contributions, as it must for a boundary term. Second, $F_{(3)}G^{ab}\phi_{ab}$ contains the Ricci tensor, so varying it produces an additional contribution that cannot be recovered from the Hessian variation alone.

The complete metric field equations can be written as
\begin{equation}\label{eq:metriceq}
\chi G_{ab}=\frac12 T^m_{ab}+\mathcal S_{ab},
\end{equation}
where
\begin{equation}
\mathcal S_{ab}=
\frac12(f-\chi R+\mathcal J)g_{ab}
+\frac12 f_Xx_ax_b+\nabla_a\nabla_b\chi-g_{ab}\Box\chi-M_{ab}-I^Y_{ab}-I^R_{ab},
\end{equation}
and we define
\begin{equation}
\chi\equiv f_R,\quad
\mathcal J\equiv\mathcal J_0+F_{(3)}G^{cd}Y_{cd}.
\end{equation}
We adopt the conventional normalization of the matter energy-momentum tensor, $\delta S_m=-\frac12\int d^4x\sqrt{-g}\,T^m_{ab}\delta g^{ab}$, which accounts for the factor $1/2$ in Eq.~\eqref{eq:metriceq}. The tensor $\mathcal S_{ab}$ collects the geometric contributions beyond the Einstein tensor, including the effects of the nonminimal curvature coupling and the quadratic-cubic scalar interactions. Its expression follows from the complete metric variation, retaining both the algebraic contributions and those arising from variations of the connection and curvature. In particular, the term $\mathcal Jg_{ab}/2$, originating from the variation of the volume element, contributes to the effective energy density and pressure but not to the heat flux or anisotropic stress. Its inclusion is necessary for the full metric response to respect integrations by parts; for instance, a constant $Q\Box\phi$ is a total divergence and must not contribute to the bulk field equations.

\subsection{Imperfect-fluid decomposition}

To identify the fluid properties associated with these gravitational interactions, we first rearrange Eq.~\eqref{eq:metriceq} into an Einstein-like form,
\begin{equation}\label{eq:rearrangement}
G_{ab}=\frac{1}{2\chi}T^m_{ab}
+T^{\rm geom}_{ab},
\quad
T^{\rm geom}_{ab}\equiv\frac{\mathcal S_{ab}}{\chi},
\end{equation}
where $\chi\neq0$. The tensor $T^{\rm geom}_{ab}$ therefore represents the effective stress associated with the scalar and its gravitational couplings, within the chosen rearrangement of the field equations. Using the scalar congruence introduced previously, we decompose this tensor into the standard imperfect-fluid form~\cite{Faraoni:2018qdr,Nucamendi:2019uen},
\begin{equation}\label{eq:fluiddecomposition}
T^{\rm geom}_{ab}
=\rho u_au_b+p h_{ab}
+2q_{(a}u_{b)}+\pi_{ab},
\end{equation}
where $q_a$ is spatial, while $\pi_{ab}$ is symmetric, spatial, and trace-free,
\begin{equation}
u^aq_a=0,\quad
u^a\pi_{ab}=0,\quad
\pi^a{}_a=0.
\end{equation}
The four effective-fluid quantities: the energy density $\rho$, the pressure $p$, the heat flux $q_a$ and the anisotropic stress $\pi_{ab}$ follow directly from the corresponding projections
\begin{equation}\label{eq:proj}
\rho=u^au^bT_{ab}^{\rm geom},\quad p=\frac13h^{ab}T_{ab}^{\rm geom},\quad
q_a=-h_a{}^bu^cT_{bc}^{\rm geom},\quad
\pi_{ab}=T^{\rm geom}_{\langle ab\rangle}.
\end{equation}

These definitions provide a correspondence between the gravitational field equations and the effective-fluid variables. Each interaction in the action contributes to these quantities through the same metric variation, allowing its effects on energy density, pressure, heat transport, and anisotropic stress to be identified consistently. In particular, the heat flux and anisotropic stress will be the central quantities in the subsequent comparison with Eckart's constitutive equations.

The effective-fluid representation depends on the choice of the coefficient $\chi$ extracted from the Einstein tensor. We adopt $\chi=f_R$ for the representative under consideration. After the preliminary exclusion of $F_{(3),X}$, the remaining $F_{(3)}(\phi)$ is absorbed into the canonical representative of Sec.~\ref{sec:boundaryequiv}, and $\chi$ is extracted anew. All subsequent constitutive restrictions refer to this fixed canonical split. Integrations by parts preserve the bulk equations but may change the effective-fluid representation~\cite{Gallerani:2024gdy}.

\section{Contributions from each interaction}\label{sec:contributions}

The geometric stress tensor obtained in the previous section contains contributions from the curvature coupling, the scalar kinetic sector, and the quadratic and cubic Hessian interactions. Although all these terms enter the same metric field equations, they affect the effective-fluid quantities differently. We therefore examine their contributions separately, using the projections in Eq.~\eqref{eq:proj}. This allows us to identify the origin of the heat flux and anisotropic stress before combining the different sectors and imposing Eckart's constitutive relations.

\subsection{Curvature dependence}

We begin with the contribution of $f(\phi,X,R)$. Its metric variation contains terms proportional to $g_{ab}$ and $x_ax_b$, together with derivatives of $\chi=f_R$. The first two contribute only to the effective energy density and isotropic pressure, whereas derivatives of $\chi$ can also generate heat flux and anisotropic stress. Applying the projections in Eq.~\eqref{eq:proj} to this part of the geometric stress tensor gives
\begin{subequations}\label{eq:f-fluid}
\begin{align}
\chi\rho_f&=Xf_X+\frac12(\chi R-f)+\D^2\chi-\theta\dot\chi,\label{eq:rhof}\\
\chi p_f&=\frac12(f-\chi R)+\ddot\chi-a^c\D_c\chi-\frac23\D^2\chi+\frac23\theta\dot\chi,\label{eq:pf}\\
\chi q_a^f&=\mathcal Q_a^f,\label{eq:qf}\\
\chi\pi^f_{ab}&=\D_{\langle a}\D_{b\rangle}\chi-\dot\chi\sigma_{ab}.\label{eq:pif}
\end{align}
\end{subequations}
Here and throughout this section, the superscript identifies the contribution of a particular interaction to the geometric fluid, rather than a separately conserved component.

The heat flux follows from the mixed projection of the second derivatives of $\chi$. To express it in terms of the scalar-frame kinematic quantities, we use the chain rule for $\chi(\phi,X,R)$, together with $\D_a\phi=0$, $\D_aX=-s^2a_a$, and $\dot X=sd$. After separating the terms proportional to the acceleration, we obtain
\begin{equation}\label{eq:qa-accelerated}
\begin{aligned}
\mathcal Q_a^f&=-\mathcal C_f a_a
-s\chi_X\big(\D_a d+sK_a{}^ba_b\big)\\
&-\chi_R\big(\D_a\dot R-K_a{}^b\D_bR\big)
+\mathcal V_f\D_aR,
\end{aligned}
\end{equation}
where
\begin{equation}\label{eq:Cfdef}
\begin{aligned}
\mathcal C_f={}&s\chi_\phi+s^3\chi_{\phi X}
-sd(\chi_X+s^2\chi_{XX})
-s^2\chi_{XR}\dot R,\\
\mathcal V_f={}&s\chi_{\phi R}
-sd\chi_{XR}-\chi_{RR}\dot R.
\end{aligned}
\end{equation}

It is possible to see that only the first term is necessarily aligned with $a_a$. The remaining terms involve spatial derivatives of $d$ and $R$, as well as contractions with $K_{ab}$, which generally define independent spatial directions. Similarly, the anisotropic stress in Eq.~\eqref{eq:pif} contains a trace-free spatial Hessian that need not be proportional to the shear. These structures provide the first restrictions to be examined in the Eckart matching. In the particular case $\chi=F_{(2)}(\phi)$, the additional heat-flow directions disappear, leaving
\begin{equation}
q_a^f=-\frac{sF_{(2),\phi}}{F_{(2)}}a_a.
\end{equation}

\subsection{Kinetic terms and braiding}

We next consider the contributions of the scalar kinetic sector and kinetic gravity braiding. Although the general function $f(\phi,X,R)$ already includes kinetic and potential terms, their contribution can be isolated by writing
\begin{equation}
f(\phi,X,R)=\mathcal F(\phi,X,R)+P(\phi,X),
\end{equation}
where $\mathcal F$ retains arbitrary curvature dependence and $P$ represents a general k-essence Lagrangian, including a possible scalar potential. This separation does not require $f$ to be linear in $R$, and the coefficient of the Einstein tensor remains $\chi=f_R=\mathcal F_R$.

Since $P$ depends only on $\phi$ and $X$, its metric variation produces terms proportional to $g_{ab}$ and $x_ax_b$. Using $x_a=su_a$, their mixed and trace-free spatial projections vanish, while the temporal and spatial-trace projections give
\begin{equation}
\begin{aligned}
\chi\rho_P&=XP_X-\frac12P,
&\chi p_P&=\frac12P,\\
q_a^P&=0,
&\pi_{ab}^P&=0.
\end{aligned}
\end{equation}
These expressions hold for arbitrary curvature dependence in $\mathcal F$. Thus, the purely k-essence contribution modifies the effective energy density and isotropic pressure without generating heat flux or anisotropic stress. Any additional effects associated with mixed curvature-kinetic dependence remain included in the $\mathcal F$ sector and are described by the curvature contributions derived above.

The braiding interaction $Q(\phi,X)\Box\phi$ differs from the purely kinetic sector because it depends explicitly on the scalar Hessian. Its metric variation therefore contains both an algebraic contribution, arising from the contraction $g^{ab}Y_{ab}$, and a connection contribution, arising from $\delta Y_{ab}=-x_c\delta\Gamma^c{}_{ab}$. Combining these terms and applying the effective-fluid projections yields
\begin{equation}\label{eq:Q-fluid}
\begin{aligned}
\chi\rho_Q&=XQ_\phi+sX\theta Q_X,
&\chi p_Q&=XQ_\phi-XQ_Xd,\\
\chi q_a^Q&=-sXQ_Xa_a,
&\chi\pi_{ab}^Q&=0.
\end{aligned}
\end{equation}
These results are also valid for general $f(\phi,X,R)$, with $\chi=f_R$, since the variation of the braiding interaction can be evaluated independently of the curvature sector.

The structure of the heat flux follows from the dependence of $Q$ on $\phi$ and $X$. Indeed, using the scalar-frame identities $\D_a\phi=0$ and $\D_aX=-s^2a_a$, we obtain
\begin{equation}
\D_aQ=Q_X\D_aX=-s^2Q_Xa_a.
\end{equation}
Consequently, the spatial variation of $Q$ is necessarily aligned with the acceleration, and the resulting heat flux contains no additional spatial direction. The absence of anisotropic stress follows from the cancellation of the trace-free contributions in the complete metric variation. As a consistency check, when $Q$ is constant, the interaction $Q\Box\phi$ reduces to a total divergence and makes no contribution to the bulk field equations.

Thus, these results show that neither a general k-essence interaction nor kinetic gravity braiding introduces tensorial structures incompatible with the heat-and-shear Eckart matching. While $P(\phi,X)$ contributes only to the effective energy density and pressure, $Q(\phi,X)\Box\phi$ also generates an acceleration-directed heat flux, without producing anisotropic stress. The restrictions on the gravitational action must therefore arise from the remaining interactions or from their combined contributions. The complete effective-fluid quantities for the combined curvature, kinetic and braiding sectors are independently derived in Appendix~\ref{app:FPQprojection}, providing a consistency check of the individual contributions obtained above.

\subsection{Higher-derivative scalar interactions}

We now consider the quadratic and cubic Hessian operators. Their metric variation is more complex because the metric appears both in the contractions of $Y_{ab}=\nabla_a\nabla_b\phi$ and in the connection defining the Hessian itself. Both contributions must be retained to determine the effective-fluid quantities.

To organize the calculation, we express $\mathcal J_0$ in terms of the scalar contractions $(t,U,V,S,W,Z_3)$ introduced in Sec.~\ref{sec:DHOSTbasis}. Defining $j_v\equiv\partial\mathcal J_0/\partial v$ for each contraction, straightforward differentiation gives
\begin{align}
j_t={}&Q+2A_2t+A_3S+3B_1t^2+B_2U+2B_4tS+B_5W+B_9S^2,\notag\\
j_U={}&A_1+B_2t+B_6S,\quad j_V=B_3,\notag\\
j_S={}&A_3t+2A_5S+B_4t^2+B_6U+B_8W+2B_9tS+3B_{10}S^2,\notag\\
j_W={}&A_4+B_5t+B_8S,\quad j_{Z_3}=B_7.\label{eq:js}
\end{align}

These derivatives identify how each operator enters the metric response. In particular, the Hessian momentum is obtained by differentiating $\mathcal J_0$ with respect to $Y_{ab}$. Applying the chain rule yields
\begin{align}
\Mom_0^{ab}={}&j_tg^{ab}+2j_UY^{ab}
+3j_V(Y^2)^{ab}+j_Sx^ax^b\notag\\
&+2j_Wx^{(a}v^{b)}
+j_{Z_3}\big(2x^{(a}w^{b)}+v^av^b\big).
\label{eq:momentum}
\end{align}
Here $v_a=Y_{ab}x^b$ and $w_a=Y_{ab}v^b$. The final term includes contributions from differentiating all three Hessian factors in $Z_3$, including the middle factor that produces $v^av^b$.

The algebraic part of the metric variation is obtained by differentiating the explicit inverse-metric contractions while keeping $x_a$ and $Y_{ab}$ fixed. For the heat flux, only its mixed projection is required,
\begin{equation}
\mathscr A_a\equiv h_a{}^bu^cM^{(0)}_{bc}.
\end{equation}
The terms arising solely from the $X$-dependence of the coupling functions are proportional to $x_ax_b$ and therefore vanish in this projection. Using the Hessian decomposition~\eqref{eq:hesskin}, the remaining terms give (see Appendix~\ref{app:derivationInvariants})
\begin{align}
\mathscr A_a&=s j_t a_a
+2j_U\big(sd\,a_a+s^2K_a{}^ba_b\big)+3j_V\big(sd^2a_a+s^2dK_a{}^ba_b
-s^3a^2a_a+s^3(K^2)_a{}^ba_b\big)\notag\\
&-s^3j_Sa_a
-j_W\big(2s^3d\,a_a+s^4K_a{}^ba_b\big)+j_{Z_3}\big(-3s^3d^2a_a+2s^5a^2a_a
-2s^4dK_a{}^ba_b-s^5(K^2)_a{}^ba_b\big).
\label{eq:Aexplicit}
\end{align}
Here $(K^2)_{ab}=K_a{}^cK_{cb}$. The result contains both acceleration-directed contributions and terms involving $K_a{}^ba_b$ and $(K^2)_a{}^ba_b$, which are not generally parallel to the acceleration.

Table~\ref{tab:jmap} summarizes the contribution of each quadratic and cubic operator to the coefficients in Eq.~\eqref{eq:js}. It provides a way to identify which interactions can contribute to the same tensorial structure. For example, $A_1$ enters through $j_U$, whereas $A_2$ enters through $j_t$. These contributions cannot yet be interpreted as complete heat currents, since the same operators also produce connection-variation terms that may reinforce or cancel their algebraic response.

\begin{table}[htbp]
\caption{Contributions of the individual scalar-Hessian interactions to the coefficients $j_v=\partial\mathcal J_0/\partial v$. Together with Eqs.~\eqref{eq:momentum}-\eqref{eq:allfluid}, these coefficients determine the corresponding effective-fluid contributions. A dash denotes zero. The entries are obtained before imposing Eckart matching or degeneracy conditions.}
\label{tab:jmap}
\scriptsize
\begin{tabular*}{0.91\linewidth}{@{\extracolsep{\fill}}lcccccc@{}}
\toprule
Interaction & $j_t$ & $j_U$ & $j_V$ & $j_S$ & $j_W$ & $j_{Z_3}$ \\
\midrule
$Q t$ & $Q$ & - & - & - & - & - \\
\addlinespace[3pt]
$A_1L_1^{(2)}$ & - & $A_1$ & - & - & - & - \\
$A_2L_2^{(2)}$ & $2A_2t$ & - & - & - & - & - \\
$A_3L_3^{(2)}$ & $A_3S$ & - & - & $A_3t$ & - & - \\
$A_4L_4^{(2)}$ & - & - & - & - & $A_4$ & - \\
$A_5L_5^{(2)}$ & - & - & - & $2A_5S$ & - & - \\
\addlinespace[3pt]
$B_1L_1^{(3)}$ & $3B_1t^2$ & - & - & - & - & - \\
$B_2L_2^{(3)}$ & $B_2U$ & $B_2t$ & - & - & - & - \\
$B_3L_3^{(3)}$ & - & - & $B_3$ & - & - & - \\
$B_4L_4^{(3)}$ & $2B_4tS$ & - & - & $B_4t^2$ & - & - \\
$B_5L_5^{(3)}$ & $B_5W$ & - & - & - & $B_5t$ & - \\
$B_6L_6^{(3)}$ & - & $B_6S$ & - & $B_6U$ & - & - \\
$B_7L_7^{(3)}$ & - & - & - & - & - & $B_7$ \\
$B_8L_8^{(3)}$ & - & - & - & $B_8W$ & $B_8S$ & - \\
$B_9L_9^{(3)}$ & $B_9S^2$ & - & - & $2B_9tS$ & - & - \\
$B_{10}L_{10}^{(3)}$ & - & - & - & $3B_{10}S^2$ & - & - \\
\bottomrule
\end{tabular*}
\end{table}

\subsection{The contribution from the connection variation}

The second part of the Hessian response originates from the variation of the connection,
$\delta Y_{ab}=-x_c\delta\Gamma^c{}_{ab}$. After integrating by parts, it is expressed in terms of derivatives of the Hessian momentum through Eq.~\eqref{eq:IY}.

To evaluate its mixed projection, we decompose the full momentum tensor into components parallel and orthogonal to the scalar congruence,
\begin{equation}\label{eq:mparts}
m=\Mom_{ab}u^au^b,\quad
m_a=h_a{}^bu^c\Mom_{bc},\quad
m_{ab}=h_a{}^ch_b{}^d\Mom_{cd},
\end{equation}
such that
\begin{equation}
\Mom_{ab}=m u_au_b-2u_{(a}m_{b)}+m_{ab}.
\end{equation}
Substitution into Eq.~\eqref{eq:IY}, followed by the mixed projection, gives
\begin{equation}\label{eq:connection}
\begin{aligned}
h_a{}^bu^c I^Y_{bc}
&=\CA_a[m,m_b,m_{bc}]\\
&=\frac{s}{2}\big(\D^bm_{ab}+ma_a-m_{ab}a^b\big)
-sK_a{}^bm_b.
\end{aligned}
\end{equation}
In obtaining this expression, the time derivatives of the spatial momentum vector cancel. The remaining spatial divergence, $\D^bm_{ab}$, contains derivatives of the Hessian and is particularly important for the subsequent matching. These contributions cannot be discarded without potentially eliminating cancellations between different higher-derivative operators.

For the quadratic-cubic sector, the momentum components follow by projecting Eq.~\eqref{eq:momentum} and using Eq.~\eqref{eq:hesskin}. The resulting expressions are
\begin{subequations}\label{eq:mexplicit}
\begin{align}
m_0={}&-j_t-2dj_U+3j_V(s^2a^2-d^2)+s^2j_S+2s^2d j_W
+(3s^2d^2-2s^4a^2)j_{Z_3},\\
(m_0)_a={}&2sj_Ua_a+3j_V(sd\,a_a+s^2K_a{}^ba_b)-s^3j_Wa_a-j_{Z_3}(2s^3d\,a_a+s^4K_a{}^ba_b),\\
(m_0)_{ab}={}&j_t h_{ab}+2sj_U K_{ab}+3s^2j_V\big[(K^2)_{ab}-a_aa_b\big]+s^4j_{Z_3}a_aa_b.
\end{align}
\end{subequations}

Together, Eqs.~\eqref{eq:Aexplicit}, \eqref{eq:connection}, and \eqref{eq:mexplicit} determine the complete quadratic-cubic contribution to the heat flux. The algebraic and connection responses must be combined before assessing its direction, since neither contribution separately represents the metric response of an operator.

The remaining effective-fluid quantities are obtained by applying the energy-density, pressure, and trace-free spatial projections to the same metric variation. Their evaluation involves the corresponding projections of $M_{ab}$ and $I^Y_{ab}$, whose explicit expressions are collected in Appendix~\ref{app:fullmetric}. This provides the complete response without repeating the lengthy tensor contractions in the main text.

\subsection{The Einstein-tensor-Hessian coupling}

The final interaction to consider is $F_{(3)}(\phi,X)G^{ab}Y_{ab}$. Unlike the preceding scalar-Hessian operators, it depends explicitly on the curvature tensor. Consequently, its metric response contains three contributions: the algebraic variation of its index contractions, the variation of the scalar Hessian, and the variation of the Ricci tensor.

The Hessian contribution is obtained from
\begin{equation}
\Mom_{(3)}^{ab}=F_{(3)}G^{ab}.
\end{equation}
To express this contribution in terms of the scalar-frame quantities, we decompose the Einstein tensor into its temporal, mixed, and spatial components relative to the four-velocity $u^a$. For this purpose, we define
\begin{equation}
G_{uu}=G_{ab}u^au^b,\quad
g_a=h_a{}^bG_{bc}u^c,\quad
G^\perp_{ab}=h_a{}^ch_b{}^dG_{cd},
\end{equation}
such that
\begin{equation}
G_{ab}=G_{uu}u_au_b-2u_{(a}g_{b)}+G^\perp_{ab}.
\end{equation}
Here, $G_{uu}$ represents the temporal projection of the Einstein tensor, $g_a$ its mixed temporal-spatial projection, and $G^\perp_{ab}$ its purely spatial part. Since the Hessian momentum associated with this interaction is proportional to the Einstein tensor, its scalar-frame components follow directly by applying the projections defined in Eq.~\eqref{eq:mparts}, yielding
\begin{equation}\label{eq:Hmom}
\big(m_{(3)},(m_{(3)})_a,(m_{(3)})_{ab}\big)
=
\big(F_{(3)}G_{uu},F_{(3)}g_a,F_{(3)}G^\perp_{ab}\big).
\end{equation}
These components determine how the curvature enters the connection contribution to the effective heat flux through Eq.~\eqref{eq:connection}. In particular, the spatial divergence of the projected Einstein tensor introduces curvature-dependent contributions that must be combined with the remaining terms of the metric variation before assessing their compatibility with Eckart's constitutive equations. Furthermore, these quantities also enter the connection contribution through Eq.~\eqref{eq:connection}, in the same manner as the quadratic-cubic momentum components.

The algebraic metric response is computed by varying the explicit inverse-metric contractions while holding $R_{ab}$ and $Y_{ab}$ fixed. Its mixed projection can be expressed using
\begin{equation}
r_a=h_a{}^bR_{bc}u^c,\quad
R_{uu}=R_{ab}u^au^b,\quad
R^\perp_{ab}=h_a{}^ch_b{}^dR_{cd}.
\end{equation}
The result is
\begin{equation}\label{eq:AH}
\begin{aligned}
\mathscr A_a^{(3)}=F_{(3)}\Bigg[
&\left(d-\frac{t}{2}\right)r_a
+sR^\perp_{ab}a^b+sK_a{}^br_b
-s\left(R_{uu}+\frac R2\right)a_a
\Bigg].
\end{aligned}
\end{equation}

The remaining contribution comes from the Ricci-tensor variation. As established in Eq.~\eqref{eq:IR}, it is determined by
\begin{equation}
E_{ab}=F_{(3)}
\left(Y_{ab}-\frac12t g_{ab}\right).
\end{equation}
After applying the mixed projection, the corresponding heat-flux contribution takes the form
\begin{equation}\label{eq:Rflux}
\mathscr R_a=\frac12h_a{}^bu^c\left[\Box E_{bc}-2\nabla_d\nabla_{(b}E_{c)}{}^d\right].
\end{equation}
The term proportional to $g_{ab}$ in Eq.~\eqref{eq:IR} does not contribute to this projection, although it remains part of the complete metric response. The covariant derivatives must also retain their original ordering, since commuting them introduces additional curvature terms. The detailed variation and its consistency checks are presented in Appendix~\ref{app:F3audit}.

An independent check follows from the contracted Bianchi identity, which gives
\begin{equation}\label{eq:Hboundary}
F_{(3)}G^{ab}Y_{ab}=\nabla_a(F_{(3)}G^{ab}x_b)-(\nabla_aF_{(3)})G^{ab}x_b.
\end{equation}
For constant $F_{(3)}$, the interaction is a total divergence and has no bulk metric response. For a general function $F_{(3)}(\phi,X)$, the same identity gives, up to a boundary term,
\begin{equation}
F_{(3)}G^{ab}Y_{ab}\doteq-F_{(3),\phi}G^{ab}x_ax_b+F_{(3),X}G^{ab}v_ax_b.
\end{equation}
This relation will be useful when examining the curvature-dependent contributions to Eckart matching in Sec.~\ref{sec:quinticobstruction}. It also emphasizes that restrictions on individual operators must be interpreted consistently under integrations by parts.

\section{The geometric fluid and the Eckart matching}\label{sec:fluidall}

Having evaluated the individual interaction sectors, we now combine them to obtain the geometric heat flux for the full action. Its numerator contains the curvature contribution, the algebraic and connection responses of the scalar Hessian, and the additional Ricci variation associated with $F_{(3)}G^{ab}Y_{ab}$. Before imposing any restriction on the coupling functions, the result is
\begin{equation}\label{eq:fluxmain}
\begin{aligned}
\chi q_a&=-\mathcal C_f a_a
-s\chi_X\big(\D_a d+sK_a{}^ba_b\big)
-\chi_R\big(\D_a\dot R-K_a{}^b\D_bR\big)
+\mathcal V_f\D_aR\\
&+\mathscr A_a
+\CA_a[m_0,(m_0)_b,(m_0)_{bc}]
+\mathscr A_a^{(3)}
+\CA_a[F_{(3)}G_{uu},F_{(3)}g_b,F_{(3)}G^\perp_{bc}]
+\mathscr R_a .
\end{aligned}
\end{equation}
The terms in Eq.~\eqref{eq:fluxmain} have been calculated in the preceding section: $\mathcal C_f$ and $\mathcal V_f$ are given in Eq.~\eqref{eq:Cfdef}; $\mathscr A_a$ in Eq.~\eqref{eq:Aexplicit}; the quadratic-cubic momentum components in Eq.~\eqref{eq:mexplicit}; and the projected connection operator $\CA_a$ in Eq.~\eqref{eq:connection}. The final three terms give the complete $F_{(3)}$ contribution, including the Ricci variation in Eq.~\eqref{eq:Rflux}. Thus the expression retains every interaction in the specified quadratic-cubic action, including possible cancellations between operators.

To see how the other fluid variables enter, we project the same metric equation twice along $u^a$, take its spatial trace, and subtract that trace from the spatial projection. In particular, the projector $h_a{}^ch_b{}^d-\tfrac13h_{ab}h^{cd}$ removes both isotropic sources and the scalar kinetic term proportional to $x_ax_b$. For brevity, let $M_{uu}=u^au^bM_{ab}$, $I^R_{uu}=u^au^bI^R_{ab}$, and $\tau=h^{ab}m_{ab}$, with analogous notation for projected spatial tensors. We obtain
\begin{subequations}\label{eq:allfluid}
\begin{align}
\begin{aligned}
\chi\rho&=Xf_X+\tfrac12(\chi R-f-\mathcal J)+\D^2\chi-\theta\dot\chi
-M_{uu}-I^R_{uu}\\
&+\tfrac12\nabla_c(smu^c)-\nabla_c(sm^c)+sK^{ab}m_{ab},
\end{aligned}\label{eq:allrho}\\[4pt]
\begin{aligned}
\chi p&=\tfrac12(f-\chi R+\mathcal J)+\ddot\chi-a^c\D_c\chi
-\tfrac23\D^2\chi+\tfrac23\theta\dot\chi\\
&-\tfrac13 h^{ab}(M_{ab}+I^R_{ab})
-\tfrac16\nabla_c(s\tau u^c)+\tfrac{s}{3}K^{ab}m_{ab},
\end{aligned}\label{eq:allp}\\[4pt]
\begin{aligned}
\chi\pi_{ab}&=\D_{\langle a}\D_{b\rangle}\chi-\dot\chi\sigma_{ab}
-M_{\langle ab\rangle}-I^R_{\langle ab\rangle}\\
&-\tfrac12(d+s\theta)m_{\langle ab\rangle}
-\tfrac{s}{2}(h_a{}^ch_b{}^d\dot m_{cd})_{\ \mathrm{TF}}
+s(K_{(a}{}^cm_{b)c})_{\ \mathrm{TF}}.
\end{aligned}\label{eq:allpi}
\end{align}
\end{subequations}
In Eq.~\eqref{eq:allpi}, the subscript (or superscript) $\mathrm{TF}$ denotes the spatial, symmetric, trace-free projection, not the four-dimensional trace-free part. For any rank-two tensor $V_{ab}$, our convention is
\begin{equation}\label{eq:TFdefinition}
(V_{ab})_{\mathrm{TF}}\equiv V_{\langle ab\rangle} =\left(h_{(a}{}^c h_{b)}{}^d-\frac13h_{ab}h^{cd}\right)V_{cd}.
\end{equation}
In particular, the time derivative $\dot m_{ab}=u^c\nabla_c m_{ab}$ is projected spatially {after} differentiation. This convention is also used for $\bigl(K_{(a}{}^cm_{b)c}\bigr)_{\mathrm{TF}}$. Together with Eq.~\eqref{eq:fluxmain}, Eqs.~\eqref{eq:allfluid} provide the complete effective-fluid response for arbitrary $f,Q,A_i,B_j,F_{(3)}$ within the action under study. The tensors $M_{ab}$, $\Mom^{ab}$, and $E^{ab}$ were defined by the metric variation, so these expressions contain no independently postulated fluid terms. In particular, $q_a$ and $\pi_{ab}$ display the spatial structures that must be tested against first-order thermodynamics.

\subsection{Constitutive relations for the geometric fluid}\label{sec:thermalquestion}

The heat flux and anisotropic stress obtained in Eqs.~\eqref{eq:fluxmain} and \eqref{eq:allpi} are fixed by the metric field equations. Their interpretation as a first-order dissipative response requires comparison with Eckart's constitutive relations~\cite{Eckart:1940te},
\begin{equation}\label{eq:thermofundamentals}
q_a=-\kappa\big(\D_a\mathcal T+\mathcal T a_a\big),\quad
\pi_{ab}=-2\eta\sigma_{ab},\quad p_{\rm visc}=-\zeta\theta.
\end{equation}
Here $\mathcal T$ denotes temperature, $\kappa$ thermal conductivity, and $\eta$ and $\zeta$ the shear and bulk viscosities. In contrast with an ordinary material fluid, these relations are not assumed when constructing the geometric stress tensor; they are conditions to be tested against it~\cite{Faraoni:2021lfc,Giusti:2021sku,Faraoni:2023hwu}. The heat-conduction and shear-viscosity relations must hold simultaneously for the same geometric stress tensor, without restricting the scalar or metric fields to a particular solution. The heat current is a spatial vector, whereas the anisotropic stress is a spatial, symmetric, trace-free tensor. A contribution incompatible with one constitutive relation cannot be removed by a contribution in the other. Nevertheless, both conditions restrict the same functions in the action. We therefore combine all operator contributions within each channel before imposing the two constitutive laws together.

One can expand the product $\kappa \D_a T$ in Eckart's original law as $\kappa D_a \mathcal T = \D_a(\kappa\mathcal T)-\mathcal T\D_a\kappa $ giving the equivalent three-term form
\begin{equation}\label{eq:eckartpotential}
q_a=-\left[\D_a(\kappa\mathcal T)-\mathcal T\D_a\kappa
+\kappa\mathcal T a_a\right].
\end{equation}
This equivalent form of Eckart's heat-conduction equation makes explicit the relations that must hold between the coefficients of the different contributions to the geometric heat current. In particular, a heat current of the form $q_a=-[\D_a(AB)-A\D_aB+Ca_a]$ admits the direct identification $\mathcal T=A$ and $\kappa=B$ provided that $C=AB$. The distinction between the two derivative contributions is therefore important, since the geometric heat current determines their combined response rather than the temperature and thermal conductivity individually. Neglecting the term $\mathcal T\D_a\kappa$ would implicitly assume a spatially uniform conductivity and could lead to different restrictions on the gravitational interactions. We therefore retain the complete constitutive equation throughout the analysis, requiring the same effective temperature and thermal conductivity to reproduce simultaneously all independent spatial-gradient contributions and the acceleration coefficient.

A special case occurs for a temperature that depends only on $(\phi,X)$, i.e $\mathcal{T}=\mathcal{T}(\phi,X)$. The kinematic identities of Sec.~\ref{sec:DHOSTbasis} give
\begin{equation}\label{eq:temperaturegradient}
\D_a\mathcal T(\phi,X)
=\mathcal T_\phi\D_a\phi+\mathcal T_X\D_aX
=-2X\mathcal T_Xa_a,
\end{equation}
thus for this case the Eckart heat-conduction force is necessarily acceleration-directed. If $\mathcal T$ depends on additional variables, such as $R$, $d=\dot s$, or $t=\Box\phi$, their spatial derivatives can also appear, but their coefficients must satisfy Eq.~\eqref{eq:eckartpotential}. We distinguish this general test from the acceleration-aligned branch below.

The shear relation can be checked directly without assigning a
viscosity in advance. At nonzero shear, contraction with $\sigma^{ab}$
fixes the only possible scalar coefficient:
\begin{equation}\label{eq:generalShearAlignment}
\eta=-\frac{\pi_{ab}\sigma^{ab}}{2\sigma_{cd}\sigma^{cd}},
\quad \pi_{ab}=-2\eta\sigma_{ab}
\quad (\sigma_{cd}\sigma^{cd}\ne0).
\end{equation}
The contraction alone does not establish the tensor equality: every
independent trace-free component must have the same coefficient. At zero
shear the anisotropic stress must vanish, and $\eta$ must extend as a
finite, regular function. The shear condition thus complements heat
conduction, which requires one common temperature and conductivity to
reproduce every spatial component of $q_a$.

We use the heat and shear channels because their tensorial constitutive structures are specified directly. An equilibrium pressure must be identified before an unambiguous bulk-viscosity coefficient can be assigned. Moreover, the existence of a formal $\mathcal T$ and $\kappa$ does not establish their positivity, an entropy current, or a causal dissipative dynamics~\cite{Eckart:1940te,Hiscock:1985zz}.

\section{Selecting scalar-tensor interactions by first-order thermodynamics}\label{sec:matching}

The complete heat current~\eqref{eq:fluxmain} and anisotropic stress~\eqref{eq:allpi} are the starting point of the selection. We impose the full three-term Eckart heat-conduction law~\eqref{eq:eckartpotential} and the shear relation $\pi_{ab}=-2\eta\sigma_{ab}$, with one and the same $\kappa$ and $\mathcal T$ in all three heat terms. Neither is identified with a coefficient in the gravitational action. In the shear test, $\eta$ may depend smoothly on the available local scalar invariants, including Hessian and curvature invariants; we do not assume $\eta=\eta(\phi,X)$. We require only that it be a finite, regular scalar coefficient on the local field-jet domain, including configurations with $\sigma_{ab}=0$: a singular factor involving inverse powers of shear would not define an Eckart constitutive law there. Thus $\pi_{ab}$ must vanish at zero shear, and at nonzero shear no scalar $\eta$ can absorb an independently oriented trace-free tensor. After the operator reduction the {direct} heat prescription fixes $\kappa=s/[2F(\phi)]$ and $\mathcal T=-H_t(\phi,X,t,S)$; we do not assert that this exhausts other local constitutive choices. The calculations are off shell and no particular spacetime symmetry is imposed.

The order of the proof follows directly from the preceding fluid projections. Explicit curvature variations can produce independent trace-free spatial curvature and mixed-Ricci terms; those must first be isolated. The many quadratic and cubic Hessian interactions, on the other hand, share the same spatial Hessian momentum and must be tested {together}, because their highest-derivative contributions can cancel. Once the shear and non-gradient heat structures have been removed, the surviving heat current can be compared term by term with Eq.~\eqref{eq:eckartpotential}. We then recover the acceleration-only branch without repeating the momentum calculation.

\subsection{Curvature obstructions}

\subsubsection{The Einstein-tensor-Hessian interaction}\label{sec:quinticobstruction}
The $F_{(3)}G^{ab}\phi_{ab}$ interaction contributes to both the Hessian and Ricci variations, through $\mathscr A_a^{(3)}$, $\CA_a$ and $\mathscr R_a$ in Eq.~\eqref{eq:fluxmain}, as well as $I^R_{\langle ab\rangle}$ in Eq.~\eqref{eq:allpi}. Unlike a pure Hessian interaction, it can generate an intrinsic-curvature anisotropic stress even for vanishing scalar shear. The contracted Bianchi identity gives, up to a divergence,
\begin{equation}\label{eq:quinticIBP}
F_{(3)}G^{ab}\phi_{ab}\doteq
-F_{(3),\phi}G^{ab}\phi_a\phi_b
+F_{(3),X}G^{ab}v_a\phi_b,
\quad v_a=\phi_{ab}\phi^b.
\end{equation}
With $v_a=\dot X u_a+s^2a_a$, this becomes
\begin{equation}\label{eq:quinticADMidentity}
F_{(3)}G^{ab}\phi_{ab}\doteq
(-s^2F_{(3),\phi}+s\dot X F_{(3),X})G_{uu}
+s^3F_{(3),X}a^iG_{ui}.
\end{equation}
Choose off-shell data with a spatially uniform time-dependent lapse and a time-independent spatial metric of constant intrinsic scalar curvature but nonzero trace-free intrinsic Ricci tensor. Then $a_i=K_{ij}=\sigma_{ij}=0$, while $\dot X$ and the trace-free intrinsic curvature can be varied independently, and
\begin{equation}\label{eq:gausstest}
 G_{uu}=\frac12({}^{(3)}R+K^2-K_{ij}K^{ij})
 \longrightarrow\frac12{}^{(3)}R.
\end{equation}
Consequently, the spatial action contains
\begin{equation}\label{eq:quinticcurv}
 \sqrt{-g}\,\frac{s}{2}
 (\dot X F_{(3),X}-sF_{(3),\phi})\,{}^{(3)}R.
\end{equation}
In unitary gauge $\phi=t$, with zero shift and spatially uniform lapse $N=N(t)$ at the chosen point, $\phi_{ij}=\phi_{0i}=0$ and $\phi_{00}=-\partial_tN/N$. Its trace-free spatial metric variation has an independently variable coefficient of ${}^{(3)}R_{\langle ij\rangle}$:
\begin{equation}\label{eq:quinticobstruct}
\left.\frac{\partial(\chi\pi_{ij})}{\partial\dot X}\right|_{{}^{(3)}R_{\langle ij\rangle}}
=-\frac{s}{2}F_{(3),X}\,{}^{(3)}R_{\langle ij\rangle}.
\end{equation}
The sign follows the Einstein-like rearrangement; the independent spatial tensor is what matters. The trace-free metric variation is taken {before} restricting the test data to $K_{ij}=0$; otherwise first variations of terms vanishing on the test configuration could be lost. The absence of a cancellation by pure Hessian operators can be seen directly in this ADM test. For $\phi=t$ and vanishing shift, every $L_i^{(2)}$ and $L_j^{(3)}$ is built algebraically from the lapse, its derivatives, and the extrinsic curvature $K_{ij}$; none contains the intrinsic spatial Ricci tensor. On the chosen data $N=N(t)$ and $K_{ij}=0$, their spatial metric variations cannot generate the two-spatial-metric-derivative structure $\dot X\,{}^{(3)}R_{\langle ij\rangle}$. By contrast, varying the explicit coefficient of ${}^{(3)}R$ in Eq.~\eqref{eq:quinticcurv} produces precisely this tensor, while $\dot X$ and the trace-free spatial curvature can be varied independently. Hence there is no compensating $A_i$ or $B_j$ contribution to the coefficient isolated in Eq.~\eqref{eq:quinticobstruct}. With the regular shear coefficient specified above, $\pi_{ij}$ must vanish at $\sigma_{ij}=0$, so the full Eckart shear law requires
\begin{equation}\label{eq:F3XnoGo}
F_{(3),X}=0.
\end{equation}
We remove the remaining $F_{(3)}(\phi)$ by the boundary-equivalent shifts of Sec.~\ref{sec:boundaryequiv}, including $f\mapsto f-XF_{(3),\phi}R$, and work henceforth in the corresponding $F_{(3)}=0$ canonical representative with $\chi=f_R$ extracted anew. The distinction between the original and canonical curvature coefficients is retained at the end of the classification.

\subsubsection{Nonlinear dependence on $R$}\label{sec:curvatureobstruction}
The curvature projections~\eqref{eq:qa-accelerated} and \eqref{eq:pif} contain the independent highest-metric-derivative terms
\begin{equation}\label{eq:Rhighestjet}
(\chi q_a)_{[\nabla\nabla R]}=-f_{RR}\D_a\dot R,
\quad
(\chi\pi_{ab})_{[\nabla\nabla R]}=f_{RR}\D_{\langle a}\D_{b\rangle}R.
\end{equation}
The second term cannot be proportional to arbitrary $\sigma_{ab}$; the Hessian operators cannot cancel its independent fourth-metric-derivative tensor. Thus the shear condition itself implies
\begin{equation}\label{eq:flinear}
f_{RR}=0,\quad f(\phi,X,R)=F_{(2)}(\phi,X)R+P(\phi,X).
\end{equation}
This does not restrict $P$. Importantly, $F_{(2),X}$ is {not} set to zero yet: its third-scalar-derivative terms can cancel against the quadratic Hessian momentum.

\subsection{Hessian response}\label{sec:principal}

Having examined the independent curvature contributions, we now investigate the restrictions arising from the combined response of the quadratic and cubic scalar-Hessian interactions. Since these interactions depend explicitly on second derivatives of the scalar field, their metric variation introduces third derivatives through the connection contribution. These terms are particularly useful for examining compatibility with Eckart's constitutive equations, since the acceleration and shear of the scalar congruence depend only on first and second derivatives of the scalar field. Consequently, by varying the third derivatives while keeping the lower-order derivatives fixed, we can examine whether the resulting contributions to the anisotropic stress remain proportional to the same shear tensor. This allows us to establish restrictions on the gravitational interactions without prescribing the functional dependence of the geometric shear viscosity in advance.

To make the independent tensorial structures explicit, we choose a comoving local inertial frame at an arbitrary spacetime event, in which $g_{ab}=\eta_{ab}$, $\Gamma^a{}_{bc}=0$, and $u^a=(1,0,0,0)$. From the scalar-Hessian decomposition~\eqref{eq:hesskin}, its temporal, mixed, and spatial components are given by $Y_{00}=-d$, $Y_{0i}=sa_i$, and $Y_{ij}=sK_{ij}$, respectively. For convenience, we introduce the notation
\begin{equation}\label{eq:thirdjetdefinitions}
Y_{ij}=Z_{ij}=sK_{ij},\quad Y_{0i}=\omega_i=sa_i,
\quad T_{abc}=\partial_{(a}Y_{bc)},\quad z_{ij}=Z_{\langle ij\rangle}.
\end{equation}
Here, $Z_{ij}$ contains both the expansion and shear of the scalar congruence, with $z_{ij}=s\sigma_{ij}$, while $\omega_i$ describes its acceleration. The completely symmetric tensor $T_{abc}$ contains the independent third-derivative data of the scalar field. At fixed metric derivatives and fixed first and second derivatives of $\phi$, its twenty independent components can be varied freely, leaving $s$, $d$, $\theta$, $a_i$, and $\sigma_{ij}$ unchanged. Notice that choosing a local inertial frame does not require the spacetime curvature to vanish, since only the connection coefficients are set to zero at the event under consideration. Contributions associated with derivatives of the connection remain fixed under the variations considered here.

We first examine the spatial projection of the complete polynomial momentum~\eqref{eq:momentum}, since its derivatives enter both the geometric heat flux and anisotropic stress through the connection variation. Using the components introduced in Eq.~\eqref{eq:thirdjetdefinitions}, we obtain
\begin{equation}\label{eq:momspatial}
\mathcal P_{ij}=j_t\delta_{ij}+2j_UZ_{ij}
+3B_3[(Z^2)_{ij}-\omega_i\omega_j]+s^2B_7\omega_i\omega_j,
\end{equation}
where $(Z^2)_{ij}=Z_i{}^kZ_{kj}$. The terms proportional to $j_S$ and $j_W$ do not contribute directly to this spatial projection, since the spatial components of the scalar-field gradient vanish in the comoving frame. Nevertheless, the dependence of the remaining coefficients on the scalar-Hessian invariants must be retained when differentiating $\mathcal P_{ij}$.

We can now isolate the contributions containing third derivatives of the scalar field in the complete heat flux~\eqref{eq:fluxmain} and anisotropic stress~\eqref{eq:allpi}. In the canonical linear-curvature representation obtained after the preceding curvature analysis, these contributions arise from both the derivatives of the Hessian momentum and the second derivatives of the nonminimal curvature coupling $F_{(2)}(\phi,X)$. In particular, the kinetic dependence of $F_{(2)}$ introduces third scalar derivatives through the second derivatives of $X$. Combining these contributions, we obtain
\begin{subequations}\label{eq:principaljets}
\begin{align}
(q_i)_{[3]}&=\frac{s}{2F_{(2)}}\partial_j\mathcal P_{ij}+\frac{sF_{(2),X}}{F_{(2)}}T_{i00},\\
(\pi_{ij})_{[3]}&=-\frac{1}{F_{(2)}}\left[\frac{s}{2}\partial_0\mathcal P_{ij}
+sF_{(2),X}T_{0ij}\right]^{\rm TF}.
\end{align}
\end{subequations}
The subscript $[3]$ denotes the contributions proportional to the third derivatives of the scalar field, including those arising through the Hessian dependence of the momentum coefficients. Accordingly, only the third-scalar-derivative parts of $\partial_j\mathcal P_{ij}$ and $\partial_0\mathcal P_{ij}$ are understood in Eq.~\eqref{eq:principaljets}, while $\mathrm{TF}$ denotes the spatial trace-free projection defined in Eq.~\eqref{eq:TFdefinition}. The lower-order contributions remain part of the complete effective-fluid quantities and will be considered in the subsequent matching. An important aspect of Eq.~\eqref{eq:principaljets} is that the curvature coupling and the scalar-Hessian interactions contribute to the same principal tensorial structures, allowing for cancellations between them. Imposing restrictions on either contribution separately could therefore exclude otherwise admissible combinations of interactions.

Let us first examine the shear constitutive relation. Since variations of $T_{abc}$ leave the shear tensor unchanged, any corresponding variation of the anisotropic stress must remain proportional to that same tensor. In other words, although the geometric shear viscosity may itself depend on higher-derivative quantities, such a dependence cannot accommodate an anisotropic stress whose tensorial direction differs from that of the shear. Motivated by the combined principal response in Eq.~\eqref{eq:principaljets}, we define
\begin{equation}\label{eq:Uprincipal}
U_{ij}=[\mathcal P_{ij}+2F_{(2),X}Z_{ij}]^{\rm TF}.
\end{equation}
The requirement that the principal anisotropic stress remain proportional to the shear implies that the change in $U_{ij}$ induced by arbitrary variations of $T_{0ab}$ must be parallel to the fixed tensor $z_{ij}$. Since the components $T_{0ij}$ can be varied independently of $T_{000}$ and $T_{00i}$, we can first examine the dependence of $U_{ij}$ on the trace-free components of the spatial Hessian while keeping its trace and the remaining Hessian components fixed.

To establish the full consequences of this requirement, we resolve $U_{ij}$ and $z_{ij}$ into their five independent trace-free components, denoted by $U_A$ and $z_A$, respectively, with $A=1,\ldots,5$. The requirement that an arbitrary variation of the shear components in the Hessian momentum produce a response parallel to $z_A$ can then be expressed as
\begin{equation}\label{eq:shearrankone}
\delta U_A=c_B\,z_A\delta z_B,
\end{equation}
where the coefficients $c_B$ may depend on the available scalar and Hessian variables. Equivalently, each derivative $\partial U_A/\partial z_B$ must be proportional to $z_A$. However, since $U_A$ is a smooth function of the Hessian components, its mixed partial derivatives must commute. Differentiating Eq.~\eqref{eq:shearrankone} with respect to $z_C$ and interchanging $B$ and $C$ therefore gives
\begin{equation}\label{eq:generalShearIntegrability}
\delta^A_Cc_B-\delta^A_Bc_C
+z_A(\partial_Cc_B-\partial_Bc_C)=0,
\end{equation}
where $\partial_C\equiv\partial/\partial z_C$. This relation considerably restricts the freedom initially allowed in Eq.~\eqref{eq:shearrankone}. For any distinct pair of indices $B,C$, we first choose an index $A$ different from both, with $z_A\ne0$. The first two terms in Eq.~\eqref{eq:generalShearIntegrability} then vanish, implying $\partial_Cc_B=\partial_Bc_C$. Taking subsequently $A=C$ and $A=B$ gives $c_B=0$ and $c_C=0$, respectively. Repeating this argument for all independent components establishes that every $c_B$ must vanish on generic shear configurations, with the result extending to the remaining configurations by continuity.

It follows that $U_{ij}$ cannot depend on any of the five independent shear amplitudes. Furthermore, variations of the remaining Hessian components, namely $Y_{00}$, $\omega_i$, and the trace of $Z_{ij}$, must also produce a response parallel to $z_{ij}$. Since $U_{ij}$ is already independent of the shear amplitudes, its derivatives with respect to these remaining variables must be parallel to every possible shear tensor. The only trace-free tensor satisfying this requirement is the zero tensor. Consequently, $U_{ij}$ is independent of all Hessian components and, since it vanishes when the Hessian itself vanishes, it must vanish identically. From Eq.~\eqref{eq:Uprincipal}, the spatial Hessian momentum must therefore take the form
\begin{equation}\label{eq:Pshearform}
\mathcal P_{ij}=\Lambda(\phi,X,Y)\delta_{ij}
-2F_{(2),X}Z_{ij}.
\end{equation}
This constitutes the general shear-principal condition. In particular, the shear constitutive relation does not require the isotropic function $\Lambda$ to be independent of the scalar Hessian. Although its derivatives do not contribute to the trace-free principal response, its spatial gradient enters the geometric heat flux through Eq.~\eqref{eq:principaljets} and may therefore be accommodated by the temperature-gradient contribution in Eckart's heat-conduction equation. This distinction is important, since imposing an acceleration-aligned heat current at this stage would introduce restrictions that do not necessarily follow from the complete first-order constitutive equations.

We now apply the general condition~\eqref{eq:Pshearform} to the explicit quadratic and cubic polynomial momentum~\eqref{eq:momspatial}. Using the coefficients introduced in Eq.~\eqref{eq:js}, the tensor defined in Eq.~\eqref{eq:Uprincipal} becomes
\begin{equation}\label{eq:independentShearTensors}
\begin{aligned}
U_{ij}=\Big[2(A_1+F_{(2),X}+B_2t+B_6S)Z_{ij}
+3B_3(Z^2)_{ij}+(s^2B_7-3B_3)\omega_i\omega_j\Big]^{\rm TF}.
\end{aligned}
\end{equation}
At fixed $(\phi,X,t,S)$, the trace of $Z_{ij}$ and the temporal component of the scalar Hessian are fixed, whereas the trace-free part of $Z_{ij}$ and the spatial vector $\omega_i$ can still be chosen independently. We can therefore examine separately the three trace-free tensorial structures appearing in Eq.~\eqref{eq:independentShearTensors}. First, setting $\omega_i=0$ and considering a generic spatial Hessian $Z_{ij}$ with three distinct eigenvalues, the tensors $(Z^2)_{ij}^{\rm TF}$ and $Z_{ij}^{\rm TF}$ are linearly independent. Their contributions cannot cancel for arbitrary shear configurations, requiring $B_3=0$. Subsequently, varying $\omega_i$ independently of $Z_{ij}$ isolates the tensor $(\omega_i\omega_j)^{\rm TF}$, whose coefficient must likewise vanish, giving $B_7=0$. The remaining coefficient multiplying $Z_{ij}^{\rm TF}$ must vanish for independent values of $t$ and $S$. Since the coupling functions depend only on $(\phi,X)$, we finally obtain
\begin{equation}\label{eq:fullShearPrincipalCoeffs}
A_1=-F_{(2),X},\quad B_2=B_3=B_6=B_7=0.
\end{equation}

These restrictions illustrate the significance of performing the constitutive matching on the complete gravitational response. While four of the cubic scalar-Hessian interactions are excluded by the shear-principal condition, the quadratic interaction associated with $A_1$ need not vanish independently. Instead, its coefficient is fixed by the kinetic dependence of the nonminimal curvature coupling, allowing their principal anisotropic-stress contributions to cancel. Interestingly, this cancellation reproduces the relation between $A_1$ and $F_{(2),X}$ encountered in the quartic Horndeski combination~\eqref{eq:quarticdictionary}, even though no Horndeski or degeneracy conditions have been imposed. The thermodynamic matching thus independently identifies a particular relation between the curvature and scalar-Hessian sectors, rather than simply excluding every interaction that contributes higher derivatives to the anisotropic stress.

Inserting the restrictions~\eqref{eq:fullShearPrincipalCoeffs} into the original action, we can separate the compensated curvature-Hessian combination $F_{(2)}R+F_{(2),X}(t^2-U)$ from the remaining scalar-Hessian interactions, which can be collected as
\begin{equation}\label{eq:fullGeneralJ}
\begin{split}
\mathcal J_{\rm rem}&=H(\phi,X,t,S)+L(\phi,X,t,S)W,\\
H&=Qt+\widehat A_2t^2+A_3tS+A_5S^2
+B_1t^3+B_4t^2S+B_9tS^2+B_{10}S^3,\\
L&=A_4+B_5t+B_8S.
\end{split}
\end{equation}
Here, $\widehat A_2\equiv A_2-F_{(2),X}$ accounts for the quadratic contribution already included in the compensated curvature-Hessian combination. All coupling functions appearing in Eq.~\eqref{eq:fullGeneralJ} depend on $(\phi,X)$, while $H_t$ and $H_S$ denote partial derivatives at fixed $(\phi,X)$ and fixed value of the other Hessian invariant. As will be shown in the subsequent analysis of the mixed-Ricci contribution to the heat flux, the complete heat-conduction law imposes the additional restriction $F_{(2),X}=0$, after which $\widehat A_2=A_2$.

It is important to emphasize that the interactions associated with $A_2,A_3,A_5,B_1,B_4,B_9,B_{10}$ have not been excluded by the principal shear analysis. Their contributions to the spatial Hessian momentum are isotropic and may therefore generate spatial heat-flow directions without introducing a corresponding independent principal anisotropic stress. This already reveals an interesting distinction between the thermal and shear responses of higher-derivative scalar-tensor gravity. Whether these interactions admit a consistent first-order thermodynamic description must consequently be decided from the complete heat-conduction equation, together with the remaining lower-order shear contributions, rather than from the alignment of the geometric heat current with the acceleration alone.

\subsection{Complete heat current before Eckart matching}\label{sec:remaining}

It is useful to derive the heat current with all the coefficients of Eq.~\eqref{eq:fullGeneralJ} still present. From the momentum components~\eqref{eq:mexplicit} one finds, for $H+LW$,
\begin{equation}\label{eq:generalMomentumH}
\begin{aligned}
j_t&=H_t+B_5W,&j_S&=H_S+B_8W,&j_W&=L,\\
m&=-j_t+s^2j_S+2s^2dL,&m_a&=-s^3La_a,&m_{ab}&=j_t h_{ab}.
\end{aligned}
\end{equation}
The algebraic and connection heat projections, Eqs.~\eqref{eq:Aexplicit} and \eqref{eq:connection}, then combine into
\begin{equation}\label{eq:HLheatidentity}
\mathscr A_a+\CA_a[m,m_b,m_{bc}]
=\frac{s}{2}\D_a(H_t+B_5W)
-\left[\frac{s^3}{2}(H_S+B_8W)+s^3dL\right]a_a.
\end{equation}
The apparent $K_a{}^ba_b$ terms from the two variations cancel exactly; the gradient of $B_5W$ remains and has not been silently suppressed. Crucially, derivatives of the Wilson functions are retained inside the spatial gradient.

The nonminimal curvature combination contributes the mixed-Ricci and extrinsic-curvature vectors obtained from the Codazzi identity
\begin{equation}\label{eq:codazziuse}
\D_a\theta-\D^bK_{ab}=-R_{ua},\quad R_{ua}=h_a{}^bR_{bc}u^c.
\end{equation} Adding its complete response to Eq.~\eqref{eq:HLheatidentity} produces
\begin{equation}\label{eq:fullHeatBeforeObstructions}
\begin{aligned}
q_a&=\frac{s}{2F_{(2)}}\D_a(H_t+B_5W)
-\frac{s}{F_{(2)}}\left[F_{(2),\phi}+s^2F_{(2),\phi X}
+\frac{s^2}{2}(H_S+B_8W)+s^2dL\right]a_a\\
&+\frac{s^2}{F_{(2)}}(F_{(2),X}+s^2F_{(2),XX})
       (K_a{}^ba_b-\theta a_a)
-\frac{s^2F_{(2),X}}{F_{(2)}}R_{ua},
\quad R_{ua}=h_a{}^bR_{bc}u^c.
\end{aligned}
\end{equation}
This is the full {post-shear-principal} heat current, before requiring an Eckart temperature-gradient representation. It shows explicitly the gradient candidates, the acceleration term, and the two independent non-gradient vectors. 

To test the mixed-Ricci term for general local temperatures, consider an off-shell configuration on a flat spatial patch with spatially uniform $s,d$ and $a_i=\theta=0$. The trace-free extrinsic curvature can be chosen freely; in Cartesian coordinates take $K_{11}=x$, $K_{22}=-x$, $K_{23}=K_{32}=y^2/2$, with the other components zero (in fixed local units). Then $t,S,W$ are spatially constant and Eq.~\eqref{eq:codazziuse} gives $R_{ui}=\D^jK_{ij}=(1,0,y)$. Equation~\eqref{eq:fullHeatBeforeObstructions} consequently reduces to $q_i=c(1,0,y)$ with spatially constant $c=-s^2F_{(2),X}/F_{(2)}$. Here $q\cdot\operatorname{curl}q=c^2$, whereas Eckart's law with $a_i=0$ gives $q_i=-\kappa\D_i\mathcal T$ and hence $q\cdot\operatorname{curl}q=0$ for any smooth local $\kappa,\mathcal T$. This contradiction, independent of the allowed temperature dependence, requires
\begin{equation}\label{eq:curvatureforces}
F_{(2),X}=0,\quad F_{(2)}=F(\phi).
\end{equation}
This simultaneously removes the $K_a{}^ba_b$ term and the $s^2F_{(2),\phi X}$ term. The full trace-free projection now simplifies: the isotropic momentum $j_th_{ab}$ cancels the trace-free algebraic $sj_tK_{ab}$ response, whereas $LW$ leaves an acceleration-squared stress. Explicitly,
\begin{equation}\label{eq:shearWgeneral}
\pi_{ab}=\frac{sF_\phi}{F}\sigma_{ab}
-\frac{s^4}{F}(A_4+B_5t+B_8S)a_{\langle a}a_{b\rangle}.
\end{equation}
Since $a_{\langle a}a_{b\rangle}$ may be nonzero at vanishing shear and $t,S$ are independent local Hessian scalars, the shear law requires
\begin{equation}\label{eq:shearonly}
A_4=B_5=B_8=0.
\end{equation}
This eliminates $LW$, including the $B_5\D_aW$ term, {by the full shear condition}, not by an unsupported claim that its gradient vanishes. Since $F_{(2),X}=0$ we henceforth set $\widehat A_2=A_2$. Collecting the conditions already established, the canonical coefficients that must vanish before the thermal integrability test are
\begin{equation}\label{eq:fullEckartZeroCoeffs}
F_{(3),X}=f_{RR}=F_{(2),X}=A_1=A_4=0,\quad
B_2=B_3=B_5=B_6=B_7=B_8=0.
\end{equation}
Here $F_{(3)}(\phi)$ has been eliminated by the stated boundary transformation; $A_2,A_3,A_5,B_1,B_4,B_9,B_{10}$ are not set to zero. The remaining action and its {complete} two constitutive projections are
\begin{equation}\label{eq:fullEckartFamily}
\mathcal L_{\rm Eckart}=F(\phi)R+P(\phi,X)+H(\phi,X,t,S),
\end{equation}
\begin{subequations}\label{eq:fullEckartFluid}
\begin{align}
q_a&=\frac{s}{2F}\D_aH_t
-\frac{s}{2F}\bigl(s^2H_S+2F_\phi\bigr)a_a,
\label{eq:fullEckartQ}\\
\pi_{ab}&=\frac{sF_\phi}{F}\sigma_{ab},
\quad \eta_{\rm geom}=-\frac{sF_\phi}{2F},
\label{eq:fullEckartPi}
\end{align}
\end{subequations}
where its two scalar derivatives contain the independent Wilson coefficients explicitly:
\begin{equation}\label{eq:allSurvivingHderivatives}
\begin{aligned}
H_t={}&Q+2A_2t+A_3S+3B_1t^2+2B_4tS+B_9S^2,\\
H_S={}&A_3t+2A_5S+B_4t^2+2B_9tS+3B_{10}S^2.
\end{aligned}
\end{equation}
Thus $A_2,B_1$ enter the temperature-gradient channel, $A_5,B_{10}$ enter the acceleration channel, and the mixed operators $A_3,B_4,B_9$ enter both. Their gradients and acceleration coefficients must nevertheless be matched with the same constitutive fields.
The result retains {every} $t,S$ operator in Eq.~\eqref{eq:fullGeneralJ}; it is the most general candidate within the original operator basis after the independent curvature, non-gradient and shear obstructions have been removed. The remaining restrictions arise not from the direction of each individual term but from the requirement that {the same} $\kappa$ and $\mathcal T$ reproduce the full heat current.

Furthermore, the surviving higher-derivative interactions introduce a qualitatively different heat-conduction mechanism from that commonly encountered in simpler scalar-tensor theories, such as Brans-Dicke gravity~\cite{Faraoni:2025alq,Faraoni:2021lfc}. In the latter, the scalar-frame kinematics restrict the spatial temperature gradient to be aligned with the acceleration, allowing the heat current to be described through a single spatial direction. In contrast, the general and non-trivial interactions present in~\eqref{eq:resummedaction} introduce a dependence on the scalar Hessian, whose spatial variations can generate temperature gradients independently of the acceleration. Consequently, the geometric fluid may sustain a nonvanishing heat current even in the absence of fluid acceleration. This enriches the first-order thermodynamic description of scalar-tensor gravity and shows the importance of retaining the complete Eckart heat-conduction equation when investigating higher-derivative theories. As it will be demonstrated, the existence of these additional heat-flow directions not only modifies the thermal properties of gravity but also provides a means of establishing further restrictions on the gravitational interactions themselves.

\subsection{Matching of heat conduction and shear viscosity}
\label{sec:fullEckart}\label{sec:reducedShear}

The previous operator selection has already enforced the two necessary
tensorial forms: no independent spatial vector remains outside the
heat-conduction sector, and no trace-free tensor remains outside the
scalar-frame shear. These restrictions are genuinely joint. For example,
prior to eliminating the $W$ sector, the shear stress contains the
independent $a_{\langle a}a_{b\rangle}$ term in
Eq.~\eqref{eq:shearWgeneral}; requiring it to vanish when
$\sigma_{ab}=0$ gives $A_4=B_5=B_8=0$, simultaneously removing the
$B_5\D_aW$ contribution to the heat current. The remaining polynomial
$H(\phi,X,t,S)$ does not change the shear coefficient. Comparison with
$\pi_{ab}=-2\eta\sigma_{ab}$ consequently fixes
\begin{equation}\label{eq:uniqueShearEta}
{\eta_{\rm geom}=-\frac{sF_\phi}{2F},\quad
\pi_{ab}=-2\eta_{\rm geom}\sigma_{ab}.}
\end{equation}
This identification is unique in the chosen Einstein-like decomposition wherever $\sigma_{ab}\ne0$ and extends regularly to zero shear. It is independent of $Q$ and of the nonlinear Hessian coefficients in $H$. The shear requirement fixes the viscous response, while the remaining heat-current equation determines whether a consistent pair $(\kappa,\mathcal T)$ exists. Both equations therefore constrain th same selected action, but in complementary ways.

The matching has now reduced the problem to a single scalar function $H(t,S;\phi,X)$. Rather than fitting a temperature and a conductivity independently to the different pieces of $q_a$, we use its already determined geometric momentum $H_t$. In fact, the exact current~\eqref{eq:fullEckartQ} can be rearranged as
\begin{equation}\label{eq:geometricThreeTerms}
q_a=\D_a(gH_t)-H_t\D_ag-g(s^2H_S+2F_\phi)a_a,
\quad g\equiv\frac{s}{2F(\phi)},
\end{equation}
and hence we use the direct identification
\begin{equation}\label{eq:conductivitySelected}
{\kappa=g=\frac{s}{2F(\phi)}>0,\quad \mathcal T=-H_t,\quad F>0,\ X>0.}
\end{equation}
This choice takes the spatial-gradient potential directly from the Hessian momentum and the conductivity prefactor from the metric response. It supplies both constitutive functions in one step, rather than inferring only their product. It is not the only possible choice, as shown in Sec.~\ref{sec:splitfreedom}. In particular, $\kappa$ is generally not spatially constant: $\kappa_X/\kappa=1/s^2$ and $\dot\kappa/\kappa=d/s+sF_\phi/F$. The exact temperature in this construction is $-H_t$, with no additional scalar-state term.

With the identification~\eqref{eq:conductivitySelected}, the right-hand
side of Eckart's heat-conduction law is
$-\kappa(\D_a\mathcal T+\mathcal T a_a)
=g\D_aH_t+gH_ta_a$.
Comparison with the geometric current~\eqref{eq:geometricThreeTerms}
shows that their spatial-gradient terms agree identically, for arbitrary
off-shell scalar and metric data. Equality of the acceleration
coefficients then gives the {single stronger condition}
\begin{equation}\label{eq:directABcondition}
{H_t+s^2H_S+2F_\phi=0.}
\end{equation}
For the fixed positive conductivity and the exact choice $\mathcal T=-H_t$, this identity is {necessary and sufficient} for the complete Eckart heat-conduction law. It must hold as an off-shell polynomial identity in the independent Hessian variables $t,S$. Differentiation implies
\begin{equation}\label{eq:directDifferentialConsequences}
H_{tt}+s^2H_{tS}=0,\quad H_{tS}+s^2H_{SS}=0,
\end{equation}
and therefore the weaker determinant relation
\begin{equation}\label{eq:generalEckartDeterminant}
{H_{tt}H_{SS}-H_{tS}^{\,2}=0.}
\end{equation}
The determinant is an algebraic {consequence}, not a replacement for~\eqref{eq:directABcondition}. In particular, it cannot prove existence of a positive direct temperature or conductivity for an arbitrary polynomial $H$.

For the quadratic-cubic polynomial~\eqref{eq:fullGeneralJ}, comparing the independent coefficients of $1,t,S,t^2,tS,S^2$ in~\eqref{eq:directABcondition} gives
\begin{equation}\label{eq:directCoefficientRestrictions}
{\begin{gathered}
Q=-2F_\phi,\quad A_3=-\frac{2A_2}{s^2},\quad
A_5=\frac{A_2}{s^4},\\
B_4=-\frac{3B_1}{s^2},\quad
B_9=\frac{3B_1}{s^4},\quad
B_{10}=-\frac{B_1}{s^6}.
\end{gathered}}
\end{equation}
Equivalently, the entire directly matched polynomial is
\begin{equation}\label{eq:rankOneKappag}
{w=t-\frac{S}{s^2}=t-\frac{S}{2X},\quad
H=-2F_\phi(\phi)t+A_2(\phi,X)w^2+B_1(\phi,X)w^3.}
\end{equation}
Indeed, $H_t+s^2H_S=-2F_\phi$ identically. For transparency, the polynomial derivatives whose determinant yields the six weaker identities are
\begin{equation}
H_{tt}=2A_2+6B_1t+2B_4S,\quad
H_{tS}=A_3+2B_4t+2B_9S,\quad
H_{SS}=2A_5+2B_9t+6B_{10}S.
\end{equation}
The six familiar polynomial Wilson identities
\begin{equation}\label{eq:sixEckartConditions}
\begin{aligned}
4A_2A_5-A_3^2&=0,\\
A_2B_9-A_3B_4+3A_5B_1&=0,\\
3A_2B_{10}-A_3B_9+A_5B_4&=0,\\
3B_1B_9-B_4^2&=0,\\
9B_1B_{10}-B_4B_9&=0,\\
3B_4B_{10}-B_9^2&=0
\end{aligned}
\end{equation}
now follow automatically from~\eqref{eq:directCoefficientRestrictions}. Within direct matching they are weaker consequences of~\eqref{eq:directABcondition}; when $\kappa$ and $\mathcal T$ are not fixed in advance they give the necessary rank condition, with local matching on the regular patches of Sec.~\ref{sec:splitfreedom}. The pair~\eqref{eq:conductivitySelected} and the exact identity~\eqref{eq:directABcondition} give the sufficient matching construction on its stated regular $F>0$, $X>0$ domain. Temperature positivity imposes the further requirement $-H_t>0$.

For example, $F=F_0e^{\lambda\phi}$, $A_2=0$, $B_1=-\beta$, $Q=-2F_\phi$, with $F_0,\lambda,\beta>0$, give
$\kappa=s/(2F)>0$ and $\mathcal T=2F_\phi+3\beta w^2>0$ on the chosen domain. The nonlinear $w^3$ interaction realizes the {full} heat law with a genuine Hessian-dependent temperature; it is not being claimed covariantly degenerate. Other choices of $\kappa$ and $\mathcal T$ are considered next.

\subsubsection{Freedom in the constitutive split}\label{sec:splitfreedom}

The rearrangement~\eqref{eq:geometricThreeTerms} comes directly from the matching, but it is not the only one. Because $\D_a\phi=0$ and $\D_aX=-2Xa_a$, the spatial gradient of any function $\tau(\phi,X)$ points along the acceleration,
\begin{equation}\label{eq:tauGauge}
\D_a(g\tau)-\tau\,\D_ag=g\,\D_a\tau=-2Xg\,\tau_X\,a_a .
\end{equation}
Adding this identity to Eq.~\eqref{eq:geometricThreeTerms} moves part of the acceleration term into the gradient term without changing the current,
\begin{equation}\label{eq:tauShift}
q_a=\D_a\big[g(H_t-\tau)\big]-(H_t-\tau)\,\D_ag-g\big(2XH_S+2F_\phi+2X\tau_X\big)a_a .
\end{equation}
Reading this expression against Eq.~\eqref{eq:eckartpotential}, with the same $\kappa\mathcal T$ in the gradient and acceleration terms, gives $\kappa=g$ and $\mathcal T=-H_t+\tau$, subject to
\begin{equation}\label{eq:tauCondition}
H_t+2XH_S+2F_\phi+2X\tau_X-\tau=0 .
\end{equation}
The direct identification~\eqref{eq:conductivitySelected} is the case $\tau=0$, for which Eq.~\eqref{eq:tauCondition} reduces to Eq.~\eqref{eq:directABcondition}.

For $H=Qt+A_2w^2+B_1w^3$ one has $H_t+2XH_S=Q$, and Eq.~\eqref{eq:tauCondition} becomes an ordinary differential equation in $X$ at fixed $\phi$,
\begin{equation}\label{eq:tauODE}
2X\tau_X-\tau=-(Q+2F_\phi),\quad
\tau=\sqrt X\left[c(\phi)-\int^X\frac{Q+2F_\phi}{2X'^{3/2}}\,\dd X'\right],
\end{equation}
which has a solution for any $Q(\phi,X)$. With the same conductivity $\kappa=s/(2F)$, the heat law is therefore satisfied for arbitrary braiding, and the relation $Q=-2F_\phi$ in Eq.~\eqref{eq:directCoefficientRestrictions} expresses the choice $\tau=0$. The simplest example is $F(\phi)R+P(\phi,X)$, for which $H=0$ and $q_a=-(sF_\phi/F)a_a$. The direct temperature $-H_t$ vanishes in this case, whereas $\tau=2F_\phi$ gives $\mathcal T=2F_\phi$ and $\kappa\mathcal T=sF_\phi/F$. The same bulk theory written with the term $-2F_\phi\Box\phi$ of Eq.~\eqref{eq:directHealthyIBP} satisfies the direct condition with $\tau=0$. More generally, the rearrangement $Q\to Q+\bar q(\phi)$, $P\to P-2X\bar q_\phi$ leaves the field equations unchanged and shifts $H_t$ by $\bar q$; it is compensated by $\tau\to\tau+\bar q$, so that $-H_t+\tau$ is unaffected while $-H_t$ alone is not.

The conductivity can be left free as well. For local $\kappa$ and $\mathcal T$ depending on $(\phi,X,t,S)$, the independence of $\D_at$, $\D_aS$ and $a_a$ turns Eckart's law into
\begin{equation}\label{eq:generalPairConditions}
\kappa\,\mathcal T_t=-gH_{tt},\quad
\kappa\,\mathcal T_S=-gH_{tS},\quad
\kappa\big(2X\mathcal T_X-\mathcal T\big)=-g\big(2XH_{tX}+2XH_S+2F_\phi\big).
\end{equation}
Where $H_{tt}$ and $H_{tS}$ do not both vanish, the first two equations give $\mathcal T=\Phi(H_t;\phi,X)$ and $\kappa=-g/\Phi'$, and the third becomes $(2X\Phi_X-\Phi)/\Phi'=2XH_S+2F_\phi$. Its left-hand side depends on $t$ and $S$ only through $H_t$, so $H_S$ must do the same, that is,
\begin{equation}\label{eq:generalRankCondition}
H_{tt}H_{SS}-H_{tS}^2=0 .
\end{equation}
On a regular patch where $(H_{tt},H_{tS})\ne(0,0)$, this condition makes the third equation a linear first-order equation for $\Phi$, admitting local solutions with $\Phi'\ne0$. On a patch where $H_{tt}=H_{tS}=0$ identically, $\mathcal T$ depends only on $(\phi,X)$; where the acceleration coefficient is nonzero, one can choose the temperature factor in the third equation to have the appropriate sign and solve for a regular conductivity. An identically vanishing current admits $\mathcal T=sC(\phi)$. The rank condition is therefore necessary, with local matching established on the regular patches just described. Smooth extension through exceptional states and positivity must be checked separately. For the quadratic-cubic polynomial, the rank condition is equivalent to the six identities~\eqref{eq:sixEckartConditions}. At fixed $(\phi,X)$, the determinant condition~\eqref{eq:sixEckartConditions} requires the quadratic and cubic terms to share a common Hessian direction, so that the nonlinear sector takes the form $C_2(\alpha t+\beta S)^2+C_3(\alpha t+\beta S)^3$. On patches where $\alpha\ne0$, the coefficient relations are $A_3=-2\lambda A_2$, $A_5=\lambda^2A_2$, $B_4=-3\lambda B_1$, $B_9=3\lambda^2B_1$, and $B_{10}=-\lambda^3B_1$, yielding
\begin{equation}\label{eq:lambdaFamily}
H=Qt+A_2(t-\lambda S)^2+B_1(t-\lambda S)^3,
\end{equation}
with $Q(\phi,X)$ and $\lambda(\phi,X)=-\beta/\alpha$ remaining arbitrary. Substitution into Eq.~\eqref{eq:generalRankCondition} verifies that the determinant vanishes identically. The complementary case $\alpha=0$ gives the acceleration-aligned family $H=Qt+A_5S^2+B_{10}S^3$, which can also be recovered as $\lambda\to\infty$ with an appropriate rescaling of $A_2$ and $B_1$. The direct prescription~\eqref{eq:conductivitySelected} imposes the stronger coefficient relations~\eqref{eq:directCoefficientRestrictions} and is therefore a particular member of this general family.

As an explicit example beyond the direct prescription, consider $\lambda=k/(2X)$ with constant $k>0$, $Q=0$, and $H=A_2(t-\lambda S)^2$. The constitutive pair
\begin{equation}\label{eq:kPair}
\kappa=\frac{s}{2F}\,X^{(k-1)/2},\quad
\mathcal T=X^{(1-k)/2}\left[\frac{2F_\phi}{k}-2A_2(t-\lambda S)\right]
\end{equation}
satisfies Eqs.~\eqref{eq:generalPairConditions}, demonstrating explicitly the freedom in the temperature-conductivity identification. The direct prescription corresponds to $\lambda=1/(2X)$, $Q=-2F_\phi$, and $\tau=0$.

Eckart's law alone therefore restricts the reduced family~\eqref{eq:fullEckartFamily} through the rank condition~\eqref{eq:generalRankCondition}, while the coefficient relations~\eqref{eq:directCoefficientRestrictions} belong to the direct normalization.

\subsection{The acceleration-aligned case as a specialization}\label{sec:accelbranch}

For comparison with earlier scalar-tensor thermodynamics, consider the distinct, purely direction-based requirement that the geometric heat current be parallel to $a_a$ for all off-shell fields, as happens when the temperature is confined to the scalar state space $(\phi,X)$. Unlike the gradient-active direct construction, independently variable $\D_at$ and $\D_aS$ are now forbidden. Equation~\eqref{eq:fullEckartQ} gives directly
\begin{equation}\label{eq:accelerationHessConstraints}
H_{tt}=H_{tS}=0
\quad\Longrightarrow\quad
A_2=A_3=B_1=B_4=B_9=0.
\end{equation}
At the highest-derivative stage these are equivalent to the eleven conditions of the acceleration-only momentum comparison, now obtained without repeating that calculation:
\begin{equation}\label{eq:principalconditions}
A_1=-F_{(2),X},\quad A_2=F_{(2),X},\quad A_3=0,\quad
B_1=B_2=B_3=B_4=B_5=B_6=B_7=B_9=0.
\end{equation}
Including the lower-order mixed-Ricci and shear restrictions gives, in the canonical $F_{(3)}=0$ representative,
\begin{equation}\label{eq:zeroall}
F_{(2),X}=0,\quad A_1=A_2=A_3=A_4=0,\quad
B_1=\cdots=B_9=0,\quad A_5,B_{10}\ \text{unrestricted}.
\end{equation} The resulting aligned action is
\begin{equation}\label{eq:selected}
\mathcal L_{\rm aligned}=F(\phi)R+P(\phi,X)+Q(\phi,X)\Box\phi
+A_5(\phi,X)S^2+B_{10}(\phi,X)S^3.
\end{equation}
Since $S=-s^2d$ and $H_S=2A_5S+3B_{10}S^2$, the {full} formula~\eqref{eq:fullEckartQ} becomes
\begin{equation}\label{eq:heatcollinear}
{q_a=-\frac1F\left[s(F_\phi+XQ_X)-s^5dA_5
+\frac32s^7d^2B_{10}\right]a_a},
\quad
\pi_{ab}=\frac{sF_\phi}{F}\sigma_{ab}.
\end{equation}
Thus the $A_5d$ and $B_{10}d^2$ contributions previously found are still present; they arise precisely from the $H_S$ term and were never omitted from the general Eckart formula. The purely directional test leaves $A_5,B_{10}$ free. If one {further} requires both $\kappa$ and $\mathcal T$ to depend only on $(\phi,X)$, the coefficient of $a_a$ in Eq.~\eqref{eq:heatcollinear} must be independent of $d$, and hence
\begin{equation}\label{eq:heatonly}
A_5=B_{10}=0\quad
\text{(additional scalar-state constitutive restriction)}.
\end{equation}
For the resulting $F(\phi)R+P+Q\Box\phi$ theory the temperature and conductivity are not separately fixed; they satisfy
\begin{equation}\label{eq:EckartStructure}
q_a=-\kappa(\mathcal T-2X\mathcal T_X)a_a,
\quad
\kappa(\mathcal T-2X\mathcal T_X)
=\frac{s}{F}(F_\phi+XQ_X).
\end{equation}
The independent shear identification is $\eta=-sF_\phi/(2F)$; no sign or entropy-production condition has been assumed in this geometric theorem.

Before the $F_{(3)}(\phi)$ boundary transformation, the acceleration-aligned conditions are equivalently expressed in the original basis as
\begin{equation}\label{eq:originalbasisconditions}
\begin{gathered}
f_{RR}=0,\quad F_{(3),X}=0,\quad
F_{(2),X}=F_{(3),\phi},\\
A_1=-F_{(3),\phi},\quad A_2=F_{(3),\phi},\quad
A_3=A_4=0,\\
B_1=\cdots=B_9=0,\quad A_5,B_{10}\ \text{free}.
\end{gathered}
\end{equation}
The conditions for the {directly matched} family must likewise be interpreted in the canonical representative, after the shifts~\eqref{eq:Hshifts}; the Einstein-like geometric-fluid decomposition is not invariant under changing the extracted Einstein coefficient. The fixed-pair selection is an off-shell statement within the written quadratic-cubic basis and on the specified scalar state domain. Special background symmetries may suppress individual obstructions, while positivity, degeneracy, and the number and stability of propagating degrees of freedom remain independent physical conditions investigated below.

\subsection{Summary of the thermodynamic selection}
\label{sec:selectiontable}

We summarize all constraints to the original action~\eqref{eq:resummedaction} in table Table~\ref{tab:selectioncriteria}.  Both branches share the same curvature and shear constraints, but differ in their heat constitutive requirements: the direct branch adopts the pair~\eqref{eq:conductivitySelected}, whereas the acceleration-aligned branch requires $q_a\propto a_a$ without imposing scalar-state closure.

\begin{table}[H]
\caption{Coefficient restrictions in the canonical curvature-Hessian basis, with $s^2=2X>0$ and $F>0$. Free coefficients are arbitrary smooth functions of $(\phi,X)$, subject to the indicated relations.}
\label{tab:selectioncriteria}
\small
\renewcommand{\arraystretch}{1.18}
\begin{tabular}{@{}p{0.265\linewidth}p{0.335\linewidth}p{0.335\linewidth}@{}}
\toprule
Coefficient / criterion 
& Direct Eckart matching 
& Acceleration-aligned heat \\
\midrule
$f_{RR},F_{(3),X},F_{(2),X}$ 
& All zero 
& All zero \\
$P(\phi,X),F(\phi)$ 
& Free, $F>0$ 
& Free, $F>0$ \\
$A_1,A_4$ 
& Both zero 
& Both zero \\
$B_2,B_3,B_5,B_6,B_7,B_8$ 
& All zero 
& All zero \\
$Q$ 
& $-2F_\phi$ 
& Free \\
$A_2,A_3,A_5$ 
& $A_2$ free; $A_3=-2A_2/s^2$, $A_5=A_2/s^4$ 
& $A_2=A_3=0$; $A_5$ free \\
$B_1,B_4,B_9,B_{10}$ 
& $B_1$ free; $B_4=-3B_1/s^2$, $B_9=3B_1/s^4$, $B_{10}=-B_1/s^6$ 
& $B_1=B_4=B_9=0$; $B_{10}$ free \\
Geometric viscosity 
& $\eta=-sF_\phi/(2F)$ 
& $\eta=-sF_\phi/(2F)$ \\
\midrule
Selected $H$ 
& $-2F_\phi t+A_2w^2+B_1w^3$, with $w=t-S/s^2$ 
& $Qt+A_5S^2+B_{10}S^3$ \\
Covariant degeneracy 
& $A_2=B_1=0$ 
& $A_5=B_{10}=0$ \\
Degenerate endpoint 
& Nonminimally coupled k-essence (up to a divergence) 
& Luminal Horndeski gravity, with $Q(\phi,X)$ free \\
\bottomrule
\end{tabular}
\end{table}

\section{Thermodynamic admissibility: temperature and geometric shear signs}\label{sec:thermalpositivity}
The six determinant identities alone do not establish the direct constitutive representation or a positive temperature. The stronger identity~\eqref{eq:directABcondition} provides the representation for the fixed pair~\eqref{eq:conductivitySelected}, and $-H_t>0$ remains a separate positivity condition. This distinction is especially important in the gradient-active case, since $t$ and $S$ contain second scalar derivatives and the corresponding $\mathcal T(\phi,X,t,S)$ is a local {extended derivative-state} potential rather than a conventional equilibrium-state temperature. We now impose the thermodynamic inequalities, in the chosen Jordan-frame representative and on a specified physical state domain, {before} asking which of the matched actions are dynamically degenerate.

For the direct branch~\eqref{eq:conductivitySelected}, temperature positivity is simply
\begin{equation}\label{eq:positiveGeneralTemperature}
{\mathcal T=-H_t>0\quad\Longleftrightarrow\quad H_t<0.}
\end{equation}
The full direct condition~\eqref{eq:directABcondition}, and not merely the sign of $H_t$, is necessary for complete Eckart matching. Using~\eqref{eq:rankOneKappag}, the actual temperature is
\begin{equation}\label{eq:HtPositivityPolynomial}
{\mathcal T=2F_\phi-2A_2w-3B_1w^2,\quad w=t-S/s^2.}
\end{equation}
If the physical state domain includes all real $w$, strict positivity for every $w$ is equivalent to
\begin{equation}\label{eq:HtAllStates}
{\begin{aligned}
&B_1<0,\quad 2F_\phi+\frac{A_2^2}{3B_1}>0,\\[-1pt]
&\hspace{14mm}\text{or}\quad B_1=A_2=0,\quad F_\phi>0.
\end{aligned}}
\end{equation}
On a bounded state interval, weaker pointwise bounds suffice. The inequalities assume the specified canonical action representative and the exact normalization $\mathcal T=-H_t$; an integration-by-parts redefinition of $H$ must not silently be interpreted as the same temperature. The six determinant identities alone do not guarantee these inequalities.

The geometric shear viscosity is unchanged by the direct heat-law normalization,
\begin{equation}\label{eq:twoSigns}
\eta_{\rm geom}=-\frac{sF_\phi}{2F},\quad
F>0:\quad \eta_{\rm geom}<0\ \Longleftrightarrow\ F_\phi>0,
\quad \eta_{\rm geom}>0\ \Longleftrightarrow\ F_\phi<0.
\end{equation}
The usual Jordan-frame scalar-tensor description often has $\mathcal T>0$ together with $\eta_{\rm geom}<0$~\cite{Faraoni:2021jri,Faraoni:2023hwu}. Neither this negative effective viscosity nor the alternative positive-viscosity sign is, by itself, a ghost test. In particular, if one insists on the {direct} choice $\kappa=g>0$, $\mathcal T=-H_t>0$ and $H_S=0$, the exact common-temperature equation yields $\mathcal T=2F_\phi>0$, hence $\eta_{\rm geom}<0$. The sign is a consequence of those additional constitutive conditions, not a universal prediction of the general branch.

The two constitutive channels are related but not interchangeable. In the direct nonlinear branch, setting $g=s/(2F)$ and $\mathcal Y=\kappa\mathcal T$ gives the exact identity
\begin{equation}\label{eq:shearHeatJointIdentity}
{\mathcal Y+2\eta_{\rm geom}=-g(2A_2w+3B_1w^2).}
\end{equation}
At the directly matched covariantly degenerate endpoint, $A_2=B_1=0$ and $\mathcal Y=-2\eta_{\rm geom}$. Away from that endpoint, the nonlinear Hessian terms contribute to the thermal product but leave the shear coefficient unchanged. If temperature is required to be positive for {all} real $w$, Eq.~\eqref{eq:HtAllStates} implies $F_\phi>0$ and therefore $\eta_{\rm geom}<0$ for $F>0$. On a restricted state interval, $\mathcal T>0$ and $\eta_{\rm geom}>0$ need not be algebraically incompatible. These are properties of the formal constitutive representation, not an entropy-production theorem.

For comparison, other constitutive assignments in the wider degenerate Horndeski subclass can realize both $\mathcal T>0,\eta_{\rm geom}<0$ and $\mathcal T>0,\eta_{\rm geom}>0$. These assignments are not part of the fixed $\kappa=g$, $\mathcal T=-H_t$ prescription. For $F_0,\lambda,\alpha,q>0$, the following two {local Eckart matches} illustrate the distinction:
\begin{equation}\label{eq:bothShearSignsExamples}
\begin{array}{c|c|c|c}
\text{Couplings}&\mathcal T&\kappa&\eta_{\rm geom}\\ \hline
F=F_0e^{\lambda\phi},\ Q=-q
&q&s\lambda/q&-s\lambda/2<0\\[3pt]
F=F_0e^{-\lambda\phi},\ Q=-\alpha X
&\alpha X&\displaystyle\frac{s(\lambda F+\alpha X)}{F\alpha X}&+s\lambda/2>0
\end{array}
\end{equation}
Here $X>0$, and the complete three-term Eckart expression is satisfied in each row. The constant $Q$ in the first row is a removable boundary contribution and illustrates a temperature normalization, not physical braiding. In the second row the spatial temperature derivative is essential to positivity of conductivity. Both examples are covariantly degenerate, but this does not certify scalar perturbative stability on an arbitrary background. The ordinary nonnegative-shear-viscosity requirement of a material-fluid entropy production law presupposes a Gibbs relation and hydrodynamic closure not supplied by the gravitational geometric-fluid construction~\cite{Eckart:1940te,Hiscock:1985zz}. Thus the actual dynamical-health tests must be imposed separately.

\section{Covariant degeneracy and dynamical stability}\label{sec:health}
The results obtained in the previous sections show that Eckart's constitutive relations provide a powerful criterion for constraining scalar-tensor interactions, substantially reducing the original operator basis without imposing any dynamical requirements. It is therefore natural to complement these thermodynamic restrictions with the conditions governing the propagating degrees of freedom. In higher-derivative scalar-tensor gravity, the DHOST framework establishes that an appropriate degeneracy of the coupled scalar-metric kinetic matrix leads to a constraint structure that eliminates the extra Ostrogradsky scalar mode~\cite{Langlois:2015cwa,Langlois:2015skt,BenAchour:2016cay,BenAchour:2016fzp,Langlois:2018dxi}. The established quadratic and cubic DHOST classifications admit nonlinear scalar-Hessian interactions through cancellations between different operators. However, many of the operators involved in such cancellations have already been constrained or excluded by Eckart matching. Our aim in this section is therefore to determine which interactions remain admissible when thermodynamic compatibility and covariant degeneracy are required simultaneously.

In this analysis, we distinguish covariant degeneracy from degeneracy restricted to scalar-unitary gauge, since the latter need not imply the same covariant constraint structure. We also distinguish degeneracy from perturbative stability, which concerns the kinetic and propagation properties of the remaining degrees of freedom. To this end, we adopt the standard arbitrary-foliation kinetic approach developed in the DHOST literature~\cite{Langlois:2015cwa,Langlois:2017mxy,Langlois:2018dxi}. These dynamical conditions complement the thermodynamic selection established above and allow us to assess how restrictive their combined application is.

\subsection{Covariant degeneracy of the restricted Hessian sector}\label{sec:covariantHeatConnection}

The quadratic and cubic DHOST classifications establish that nonlinear scalar-Hessian interactions can be dynamically admissible when their contributions to the coupled scalar-metric kinetic matrix satisfy appropriate degeneracy conditions~\cite{Langlois:2015cwa,BenAchour:2016cay,BenAchour:2016fzp,Langlois:2018dxi}. However, the preceding thermodynamic analysis has already substantially restricted the operator basis and, consequently, the possible cancellations responsible for such degeneracy. Moreover, general Eckart matching requires the Hessian of $H$ with respect to $(t,S)$ to have rank at most one, as established in Eq.~\eqref{eq:generalRankCondition}. This additional restriction considerably simplifies the application of the DHOST degeneracy conditions.

Following the standard arbitrary-foliation approach of the DHOST literature, we decompose the scalar gradient with respect to the unit normal $n^a$ of an arbitrary spacelike hypersurface as
\begin{equation}\label{eq:tiltedADMnormal}
\phi_a=A_*n_a+p_a,\quad n^ap_a=0,\quad A_*^2=s^2+p^2.
\end{equation}
The scalar-unitary gauge corresponds to $p_a=0$, whereas an arbitrary foliation allows $p_a\ne0$. Denoting the highest scalar velocity by $v=n^an^b\phi_{ab}$ and the metric velocities by $K^{(n)}_{ij}$, the velocity-dependent parts of the Hessian contractions are
\begin{equation}\label{eq:boostedVelocities}
\left.t\right|_{\rm vel}=-v+A_*K^{(n)},\quad
\left.S\right|_{\rm vel}=A_*^2v-A_*p^ip^jK^{(n)}_{ij}.
\end{equation}
Since $F$ depends only on $\phi$, the gravitational kinetic block $\mathbb G$ is nonsingular for $F\ne0$. The rank-one condition imposed by Eckart matching then ensures that the determinant of the full kinetic matrix factorizes as
\begin{equation}\label{eq:kineticDeterminantFactored}
\det\mathbb K=\det\mathbb G\,\mathscr K(A_*^2),
\end{equation}
where
\begin{equation}\label{eq:boostedADMdeterminant}
\mathscr K(A_*^2)=H_{tt}-2A_*^2H_{tS}+A_*^4H_{SS}.
\end{equation}
This factorization follows from the rank-one contribution of the nonlinear Hessian sector to the coupled scalar-metric kinetic matrix, as explicitly demonstrated in Appendix~\ref{app:kineticdegeneracy}. All derivatives of $H$ are evaluated at fixed $(\phi,X)$. Covariant degeneracy requires $\mathscr K$ to vanish independently of the chosen foliation, rather than only for a particular value of $A_*^2$. Consequently, within the thermodynamically selected family, one necessarily has
\begin{equation}\label{eq:fullCovariantHealth}
H_{tt}=H_{tS}=H_{SS}=0,
\end{equation}
which, for the quadratic and cubic polynomial interactions considered here, implies
\begin{equation}\label{eq:fullDHOSTZeroCoeffs}
A_2=A_3=A_5=B_1=B_4=B_9=B_{10}=0.
\end{equation}

This result illustrates the significance of the preceding thermodynamic selection. Although the general DHOST classifications admit nontrivial degenerate combinations of higher-derivative operators, the restrictions imposed by Eckart matching leave no such nonlinear combinations compatible with full covariant degeneracy within the family under consideration. Thermodynamic matching and dynamical degeneracy therefore act as complementary selection criteria, whose combined application is more restrictive than either requirement separately.

The implications can be seen explicitly in the directly matched family $H=-2F_\phi t+A_2w^2+B_1w^3$, for which
\begin{equation}\label{eq:directKineticFactor}
\mathscr K=(2A_2+6B_1w)\left(1+\frac{A_*^2}{2X}\right)^2,\quad
\left.\mathscr K\right|_{p=0}=4(2A_2+6B_1w).
\end{equation}
Thus, the nonlinear interactions selected by the direct Eckart prescription are generically nondegenerate even in scalar-unitary gauge. Similarly, the acceleration-aligned family gives $\mathscr K=A_*^4(2A_5+6B_{10}S)$, while the general rank-one family~\eqref{eq:lambdaFamily} yields $\mathscr K=\mathscr U_{ww}(1+\lambda A_*^2)^2$. Isolated zeros of these expressions do not establish degeneracy throughout an open state domain.

An important distinction arises for the member $\lambda=-1/(2X)$ of the general family~\eqref{eq:lambdaFamily}, corresponding to the combination $z=t+S/(2X)=s\theta$. The nonlinear Hessian sector then depends on the expansion of the scalar congruence,
\begin{equation}\label{eq:UdegenerateFamily}
H=Qt+A_2z^2+B_1z^3,\quad
\mathscr K=(2A_2+6B_1z)\frac{p^4}{s^4}.
\end{equation}
Hence, the kinetic factor vanishes identically in scalar-unitary gauge but is generically nonzero in a tilted foliation. Additionally, this branch also admits positive constitutive coefficients. For $Q=0$ and a positive constant reference value $X_0$ with the dimensions of $X$, the pair
\begin{equation}\label{eq:UdegeneratePair}
\kappa=\frac{s}{2F}\frac{X_0}{X},\qquad
\mathcal T=-\frac{X}{X_0}
\left(2F_\phi+2A_2z+3B_1z^2\right)
\end{equation}
satisfies the complete Eckart heat-conduction law off shell. Indeed, using $H_t=2A_2z+3B_1z^2$, $s^2H_S=H_t$, $\D_aX=-2Xa_a$, and $\D_aF_\phi=0$, one obtains
\begin{equation}
-\kappa(\D_a\mathcal T+\mathcal T a_a)
=\frac{s}{2F}\left[\D_aH_t-(H_t+2F_\phi)a_a\right],
\end{equation}
which coincides exactly with the geometric heat current~\eqref{eq:fullEckartQ}. In particular, choosing $A_2=0$, $B_1<0$, $F>0$, and $F_\phi<0$ yields $\kappa>0$, $\mathcal T>0$, and $\eta_{\rm geom}>0$ for all real $z$ and $X>0$. Thus, genuinely nonlinear Hessian interactions can simultaneously satisfy these three thermodynamic sign conditions while exhibiting degeneracy in scalar-unitary gauge. In contrast to the direct normalization, global positivity of the effective temperature does not necessarily require negative geometric shear viscosity.

The literature on U-degenerate scalar-tensor theories has shown that such a restricted degeneracy can be associated with an instantaneous or ``shadowy'' mode rather than an additional propagating degree of freedom~\cite{DeFelice:2018ewo,DeFelice:2022xvq}. The dynamical consistency and perturbative stability of this branch consequently require a separate analysis and cannot be decided solely from the covariant kinetic determinant or the thermodynamic sign conditions.
Our conclusions are for exact gravitational theories satisfying full covariant degeneracy. They do not exclude the possible interpretation of nonlinear thermodynamically matched interactions within a perturbative effective field theory, nor do they establish the perturbative stability of the surviving theories. Rather, they demonstrate how the independent restrictions arising from Eckart thermodynamics and the established DHOST framework jointly constrain scalar-tensor interactions.

\subsection{Horndeski intersection and its stability}
Both selections thus intersect the standard covariantly degenerate sector at
\begin{equation}\label{eq:generalHealthyIntersection}
{\mathcal L_{\rm deg}=F(\phi)R+P(\phi,X)+Q(\phi,X)\Box\phi.}
\end{equation}
This is the established luminal Horndeski subclass $G_2=P$, $G_3=-Q$, $G_4=F(\phi)$, $G_5=0$, with the standard scalar-metric constraint structure~\cite{Horndeski:1974wa,Kobayashi:2011nu,Kobayashi:2019hrl}. The two selection routes and their covariantly degenerate structural intersection can be summarized as follows:
\begin{equation}\label{eq:twoEckartPaths}
\begin{aligned}
\text{complete Eckart matching}&\ \longrightarrow\ F R+P+H(t,S)
\ \xrightarrow{\ \text{covariant degeneracy}\ }\ F R+P+Qt,\\
\text{acceleration-aligned matching}&\ \longrightarrow\ F R+P+Qt+A_5S^2+B_{10}S^3
\ \xrightarrow{\ \text{covariant degeneracy}\ }\ F R+P+Qt.
\end{aligned}
\end{equation}
The diagram records the operator-level intersection; the direct pair, with $\tau=0$, gives $Q=-2F_\phi(\phi)$, hence $Q_X=0$, while the acceleration-aligned route and the general pair of Sec.~\ref{sec:splitfreedom} leave $Q(\phi,X)$ arbitrary. The direct endpoint is boundary-equivalent to nonminimally coupled k-essence, since
\begin{align}
-2F_\phi\Box\phi
&\doteq 2\nabla_aF_\phi\nabla^a\phi=-4XF_{\phi\phi},\label{eq:directHealthyIBP}\\
{\mathcal L_{\mathrm{healthy,direct}}}
&\doteq{F(\phi)R+\widetilde P(\phi,X)},\quad
\widetilde P(\phi,X)=P(\phi,X)-4XF_{\phi\phi}(\phi).
\label{eq:directHealthyKessence}
\end{align}
It contains no genuine $Q_X$-driven kinetic braiding, though nonconstant $F$ still mixes scalar and metric perturbations. The equivalence holds for the bulk action and field equations; the geometric temperature $\mathcal T=-H_t$ remains defined in the canonical representative used for matching. Moreover, the heat flux and shear stress at the common Horndeski endpoint are precisely the scalar-fluid expressions obtain for the thermodynamics of Horndeski gravity~\cite{Giusti:2021sku,Nucamendi:2019uen}:
\begin{equation}\label{eq:covariantRemovesGradient}
q_a=-\frac{s}{F}(F_\phi+XQ_X)a_a,\quad
\pi_{ab}=\frac{sF_\phi}{F}\sigma_{ab},\quad
\eta_{\rm geom}=-\frac{sF_\phi}{2F}.
\end{equation}

\section{Thermal evolution and shear viscosity}\label{sec:thermalEvolution}
We now consider the evolution of the thermal product $\mathcal Y=\kappa\mathcal T$, which determines the heat-conduction response in the chosen Eckart prescription. Following the first-order thermodynamic interpretation of scalar-tensor gravity~\cite{Faraoni:2021jri,Giardino:2022sdv,Faraoni:2022doe,Faraoni:2023hwu}, we call $\dot{\mathcal Y}<0$ local cooling and $\dot{\mathcal Y}>0$ local heating, with $\dot{\mathcal Y}=u^a\nabla_a\mathcal Y$ and a future-directed scalar congruence. These sign conditions refer to the chosen constitutive normalization and are considered on a patch where $F>0$, $X>0$, $\kappa>0$, and $\mathcal T>0$. They characterize the evolution of the effective-fluid temperature-conductivity product, not physical entropy production. In particular, a decreasing $\mathcal Y$ need not approach zero, and a thermal zero does not necessarily identify general relativity.

\subsection{Gradient-active evolution before covariant degeneracy}\label{sec:generalThermalEvolution}
For the directly matched nonlinear family~\eqref{eq:rankOneKappag}, the thermal product is
\begin{equation}\label{eq:fullEckartY}
\mathcal Y=\kappa\mathcal T
=\frac{s}{2F}\bigl(2F_\phi-2A_2w-3B_1w^2\bigr),
\quad w=t-\frac{S}{s^2}.
\end{equation}
The effective temperature $\mathcal T=-H_t$ depends on the extended derivative state. Writing $g=s/(2F)$ and
\begin{equation}\label{eq:thermalSlope}
\mathcal K\equiv 2A_2+6B_1w=-\mathcal T_w,
\end{equation}
its spatial gradient and the corresponding heat current are
\begin{equation}\label{eq:gradientActiveForce}
\begin{aligned}
\D_a\mathcal T&=-s^2\mathcal T_X\big|_w\,a_a-\mathcal K\D_aw,\\
q_a&=g\mathcal K\D_aw-g\bigl(\mathcal T-s^2\mathcal T_X\big|_w\bigr)a_a,
\end{aligned}
\end{equation}
where $\mathcal T_X|_w=-2A_{2,X}w-3B_{1,X}w^2$. Thus $\D_aw$ can supply a heat-flow direction independent of acceleration. The kinematic relations $t=d+s\theta$, $S=-s^2d$ give
\begin{equation}\label{eq:wScalarEvolution}
w=2d+s\theta,\quad \dot w=2\dot d+d\theta+s\dot\theta.
\end{equation}
Using $\dot\phi=-s$, $\dot X=sd$, and $\dot g/g=d/s+sF_\phi/F$, the evolution follows directly from Eq.~\eqref{eq:fullEckartY}:
\begin{equation}\label{eq:fullEckartEvolution}
\begin{aligned}
\dot{\mathcal Y}={}&\mathcal Y\left(\frac d s+\frac{sF_\phi}{F}\right)
+g\left[-s\mathcal T_\phi\big|_{X,w}
+sd\mathcal T_X\big|_{\phi,w}-\mathcal K\dot w\right],\\
\mathcal T_\phi\big|_{X,w}={}&2F_{\phi\phi}-2A_{2,\phi}w-3B_{1,\phi}w^2,\\
\mathcal T_X\big|_{\phi,w}={}&-2A_{2,X}w-3B_{1,X}w^2.
\end{aligned}
\end{equation}
For a compact statement of the cooling and heating conditions, denote by
\begin{equation}\label{eq:gradientHeatingThreshold}
\mathscr R\equiv\mathcal Y\left(\frac d s+\frac{sF_\phi}{F}\right)
+g\left[-s\mathcal T_\phi\big|_{X,w}
+sd\mathcal T_X\big|_{\phi,w}\right]
\end{equation}
the part of $\dot{\mathcal Y}$ independent of $\dot w$. Since $g>0$,
\begin{equation}\label{eq:generalEckartHeatingCooling}
\begin{aligned}
\dot{\mathcal Y}<0\ \ &(\text{cooling})
\quad\Longleftrightarrow\quad \mathcal K\dot w>\frac{\mathscr R}{g},\\
\dot{\mathcal Y}>0\ \ &(\text{heating})
\quad\Longleftrightarrow\quad \mathcal K\dot w<\frac{\mathscr R}{g}.
\end{aligned}
\end{equation}
The terms involving $\dot d$ and $\dot\theta$ show why expansion alone does not determine the sign of $\dot{\mathcal Y}$ before degeneracy. In this family $\mathcal Y=0$ is the state-dependent condition $2F_\phi=2A_2w+3B_1w^2$; it need not require $F_\phi=0$ or the disappearance of the nonlinear interactions. At such a point $q_a=-g\D_a\mathcal T$ and $\dot{\mathcal Y}=g\dot{\mathcal T}$ can remain nonzero. The local sign of $\dot{\mathcal Y}$ therefore does not establish relaxation to a GR state. Moreover, this nonlinear family is generically nondegenerate, as shown in Sec.~\ref{sec:health}, so its thermal evolution is a formal off-shell diagnostic rather than the evolution of an automatically healthy theory.

The shear viscosity, unlike the thermal product, is independent of the choice of constitutive pair within the reduced family. From $\eta_{\rm geom}=-sF_\phi/(2F)$, $\D_as=-sa_a$ and $\dot\phi=-s$, we obtain
\begin{equation}\label{eq:shearEtaEvolution}
\D_a\eta_{\rm geom}=-\eta_{\rm geom}a_a,\quad
\dot\eta_{\rm geom}=\frac{s^2}{2F}\left(F_{\phi\phi}-\frac{F_\phi^2}{F}\right)-\frac{dF_\phi}{2F}.
\end{equation}
The corresponding projected evolution of the anisotropic stress is
\begin{equation}\label{eq:piShearEvolution}
\dot\pi_{\langle ab\rangle}=-2\dot\eta_{\rm geom}\sigma_{ab}-2\eta_{\rm geom}\dot\sigma_{\langle ab\rangle}.
\end{equation}
At the directly matched covariantly degenerate endpoint, $\mathcal Y=-2\eta_{\rm geom}$ and therefore $\dot{\mathcal Y}=-2\dot\eta_{\rm geom}$. Before degeneracy, the nonlinear terms in Eq.~\eqref{eq:shearHeatJointIdentity} distinguish their evolution. The sign of $\dot\eta_{\rm geom}$ alone does not determine the evolution of the shear stress or establish a GR limit.

\section{Field-equation equivalence and scope}\label{sec:scope}\label{sec:boundaryequiv}
Equation~\eqref{eq:Hboundary} follows from $\nabla_aG^{ab}=0$ and the contracted Bianchi identity.
A constant $F_{(3)}$ is therefore a boundary operator. If $F_{(3)}=F_{(3)}(\phi)$, the identity $G^{ab}\phi_a\phi_b=t^2-U+XR-\nabla_aJ^a$, $J^a=\phi^at-\phi_b\phi^{ab}$, allows a $F_{(3)}=0$ representative with coefficient shifts
\begin{equation}\label{eq:Hshifts}
\begin{aligned}
F_{(2)}&\mapsto F_{(2)}-XF_{(3),\phi},&
A_1&\mapsto A_1+F_{(3),\phi},\\
A_2&\mapsto A_2-F_{(3),\phi},&
Q&\mapsto Q+3XF_{(3),\phi\phi},\\
P&\mapsto P-2X^2F_{(3),\phi\phi\phi}.
\end{aligned}
\end{equation}
Bulk metric equations are identical under this boundary equivalence. To compare effective fluids one must keep {the same} extracted Einstein coefficient $\chi$; otherwise the unnormalized geometric source $\mathcal S_{ab}$ changes by $(\widehat\chi-\chi)G_{ab}$ and its normalization changes as well. We impose the normal form after the preliminary $F_{(3),X}$ test and before the remaining constitutive tests, assuming its reduced $F_{(2)}\ne0$.

Within the stated operator basis and canonical Jordan-frame Einstein split, the independent curvature, shear and heat-vector tests select an action polynomial $H(\phi,X,t,S)$. The direct matching prescription $\kappa=s/(2F)>0$, $\mathcal T=-H_t$ is fully characterized by the stronger off-shell identity~\eqref{eq:directABcondition}. Its six determinant relations are weaker than that identity, but they are necessary when $\kappa$ and $\mathcal T$ are left free, with local sufficiency on the regular patches described in Sec.~\ref{sec:splitfreedom}. Positive temperature requires $H_t<0$ on the chosen domain. The relation $Q=-2F_\phi$ belongs to the direct normalization; other pairs allow any $Q$, and their positivity has to be checked separately.

The effective scalar temperature is a derivative-state geometric-fluid potential, not a physical equilibrium temperature obtained from a Gibbs relation or entropy current. Constitutive matching does not establish entropy production or causal thermal dynamics. The scalar-flow evolution of this gradient-active potential depends on $\dot d$ and $\dot\theta$ as shown in Sec.~\ref{sec:generalThermalEvolution}. Covariant kinetic degeneracy removes its genuinely nonlinear $H$ interactions; in the directly matched subfamily this leaves $F(\phi)R+P(\phi,X)-2F_\phi(\phi)\Box\phi$, or equivalently the nonminimally coupled k-essence action in Eq.~\eqref{eq:directHealthyKessence}. The more general, possibly genuinely $Q_X$-braided Horndeski form $F(\phi)R+P+Q\Box\phi$ is the degenerate endpoint of both the acceleration-aligned branch and the general pairs of Sec.~\ref{sec:splitfreedom}. Boundaries and integrations by parts change the assigned geometric-fluid representative, so all constitutive choices and sign statements refer to the fixed one. This dependence on the representation is also relevant to invariant formulations: a Jordan-like invariant construction can retain an imperfect-fluid thermodynamic description~\cite{Pereira:2026ccq}.

\section{Conclusions}\label{sec:conclusions}

In this work, we have investigated the extent to which first-order thermodynamics can constrain the structure of scalar-tensor gravity beyond the restrictions previously established within particular classes of gravitational theories. To this end, we considered a broad action containing nonlinear curvature dependence, an Einstein-tensor-Hessian coupling, and the complete set of independent quadratic and cubic scalar-Hessian interactions, without imposing degeneracy conditions or relations between the coupling functions. From the complete metric variation, we derived the general expressions for the effective energy density, pressure, heat flux, and anisotropic stress, given by Eqs.~\eqref{eq:fluxmain} and~\eqref{eq:allfluid}. These expressions constitute a result of independent interest, providing a covariant effective-fluid description of higher-derivative scalar-tensor gravity that is not restricted to Horndeski or DHOST theories, particular solutions, or spacetime symmetries.

The simultaneous matching of the geometric heat flux and anisotropic stress with Eckart's constitutive equations proved to be restrictive. By considering the combined response of the different interactions, we established which contributions can compensate for one another and which remain incompatible with first-order thermodynamics. In particular, even when the scalar-Hessian interactions compensate for the principal shear contributions arising from the kinetic dependence of the nonminimal curvature coupling, the remaining mixed-Ricci contribution to the heat flux cannot be accommodated by Eckart's heat-conduction law. Together with the restrictions arising from the complete shear response, this substantially reduces the original gravitational action to the family~\eqref{eq:fullEckartFamily}, in the canonical representation described in Sec.~\ref{sec:scope}. Moreover, despite the presence of surviving quadratic and cubic interactions, the geometric shear viscosity is determined entirely by the nonminimal curvature coupling.

A central result of this work emerges from retaining the complete Eckart heat-conduction equation rather than requiring the geometric heat current to be directed along the scalar-frame acceleration. The existence of a common effective temperature and thermal conductivity imposes the integrability condition~\eqref{eq:generalRankCondition}, which translates directly into the relations~\eqref{eq:sixEckartConditions} between the quadratic and cubic coupling functions. Consequently, the surviving nonlinear interactions must depend on a common scalar-Hessian combination, as expressed by Eq.~\eqref{eq:lambdaFamily} and its pure-$S$ counterpart. This restriction is particularly relevant because it follows from the existence of a local constitutive representation, rather than from prescribing the temperature and conductivity in advance. Particular identifications, such as the direct prescription~\eqref{eq:conductivitySelected}, can impose further restrictions, allowing us to distinguish the conditions intrinsic to Eckart matching from those associated with the choice of thermodynamic quantities.

The surviving nonlinear interactions also reveal new features in the thermal description of scalar-tensor gravity. Their contribution to the effective temperature introduces spatial gradients that can sustain a geometric heat current independently of the fluid acceleration. For the directly matched family, we derived the corresponding thermal evolution~\eqref{eq:fullEckartEvolution}, together with the local conditions governing heating and cooling, and the evolution of the geometric shear viscosity and anisotropic stress. The relation~\eqref{eq:shearHeatJointIdentity} shows that the nonlinear interactions modify the thermal response without altering the shear viscosity, allowing these two thermodynamic quantities to encode different properties of the gravitational action. Furthermore, a vanishing thermal product does not necessarily imply a vanishing heat flux or relaxation towards general relativity. We have also established that positive temperature and conductivity need not require negative geometric shear viscosity, as demonstrated by the constitutive representations discussed in Sec.~\ref{sec:thermalpositivity} and the nonlinear example~\eqref{eq:UdegeneratePair}.

Perhaps the most interesting implication of the thermodynamic restrictions arises when they are complemented with covariant degeneracy. The established DHOST classifications allow nonlinear scalar-Hessian interactions through cancellations in the coupled scalar-metric kinetic matrix. However, the thermodynamic selection considerably restricts the interactions available for such cancellations. In particular, the rank condition derived independently from Eckart's heat-conduction equation leads directly to the factorization of the kinetic determinant~\eqref{eq:kineticDeterminantFactored}. Requiring degeneracy in an arbitrary foliation then excludes all the remaining nonlinear Hessian interactions through the restrictions~\eqref{eq:fullDHOSTZeroCoeffs}. Thus, although neither first-order thermodynamic matching nor covariant degeneracy separately excludes nonlinear scalar-Hessian interactions, their combined application does so within the class of theories considered here. This establishes a direct connection between the integrability of the geometric heat current and the kinetic structure of the gravitational theory.

The intersection of these two requirements recovers the established luminal Horndeski subclass~\eqref{eq:generalHealthyIntersection}, while the direct thermodynamic prescription leads to nonminimally coupled k-essence~\eqref{eq:directHealthyKessence}. The significance of this result lies not in the identification of a new Horndeski subclass, but in the independent physical route through which it is recovered. Neither the Horndeski structure nor luminal gravitational-wave propagation was imposed on the original action. Interestingly, relaxing covariant degeneracy to degeneracy in the scalar-unitary gauge also allows a nonlinear thermodynamically compatible family, given by Eq.~\eqref{eq:UdegenerateFamily}, which admits positive temperature, conductivity, and shear viscosity. This provides a further setting in which to investigate the relation between the thermodynamic properties of gravity and the constraint structure of higher-order scalar-tensor theories.

Overall, our results show that the restrictive power of first-order thermodynamics extends considerably beyond the identification of effective gravitational temperature and transport coefficients. Within the operator basis and effective-fluid representation considered here, Eckart's constitutive equations constrain the curvature dependence of the gravitational action, establish relations between independent higher-derivative interactions, and restrict the cancellations responsible for dynamical degeneracy. While thermodynamic compatibility cannot replace the requirements of perturbative stability and observational viability, it provides an independent criterion that can be applied before such requirements are imposed. This suggests that the thermodynamic description of gravity may play a more fundamental role in the construction and classification of gravitational theories than previously considered. It would therefore be interesting to investigate whether analogous connections between constitutive integrability and gravitational kinetic structure emerge in theories with multiple scalar degrees of freedom, more general gravitational interactions, or thermodynamic descriptions extending beyond Eckart's first-order formulation~\cite{Banerjee:2026ulx}.

\begin{acknowledgments}
We thank Jess Rutschi for helpful comments. This work was supported by Fundação para a Ciência e a Tecnologia (FCT) through national funds under the research grant UID/04434/2025 (DOI 10.54499/UID/04434/2025).
\end{acknowledgments}

\bibliographystyle{apsrev4-2}
\bibliography{biblio}

\appendix

\section{Quadratic and cubic scalar-Hessian operators}\label{app:basisops}
The following standard contractions fix our operator labels and remove any ambiguity about the coefficients used in the main text. They are not subject to matching or degeneracy assumptions.
\subsection{The five quadratic operators}
The conventional contractions are
\begin{align}
L_1^{(2)}&=\phi_{ab}\phi^{ab}, &
L_2^{(2)}&=(\Box\phi)^2,\notag\\
L_3^{(2)}&=(\Box\phi)\,\phi^a\phi_{ab}\phi^b, &
L_4^{(2)}&=\phi^a\phi_{ab}\phi^{bc}\phi_c,\notag\\
L_5^{(2)}&=(\phi^a\phi_{ab}\phi^b)^2.
\label{eq:quadraticstandard}
\end{align}
The operators are independent on generic field configurations, and their notation follows Refs.~\cite{BenAchour:2016cay,Langlois:2018dxi}.

\subsection{The ten cubic operators}
The standard cubic basis is
\begin{align}
L_1^{(3)}={}&(\Box\phi)^3,\notag\\
L_2^{(3)}={}&(\Box\phi)\,\phi_{ab}\phi^{ab},\notag\\
L_3^{(3)}={}&\phi_a{}^b\phi_b{}^c\phi_c{}^a,\notag\\
L_4^{(3)}={}&(\Box\phi)^2\,\phi^a\phi_{ab}\phi^b,\notag\\
L_5^{(3)}={}&(\Box\phi)\,\phi^a\phi_{ab}\phi^{bc}\phi_c,\notag\\
L_6^{(3)}={}&(\phi_{ab}\phi^{ab})\,\phi^c\phi_{cd}\phi^d,\notag\\
L_7^{(3)}={}&\phi^a\phi_{ab}\phi^{bc}\phi_{cd}\phi^d,\notag\\
L_8^{(3)}={}&(\phi^a\phi_{ab}\phi^b)\,
                   (\phi^c\phi_{cd}\phi^{de}\phi_e),\notag\\
L_9^{(3)}={}&(\Box\phi)\,(\phi^a\phi_{ab}\phi^b)^2,\notag\\
L_{10}^{(3)}={}&(\phi^a\phi_{ab}\phi^b)^3.
\label{eq:cubicstandard}
\end{align}
These are the standard cubic DHOST basis operators \cite{BenAchour:2016fzp}.

\section{An independent covariant variation of the Einstein-tensor-Hessian interaction}\label{app:F3audit}
We give the full metric variation of the interaction
\begin{equation}\label{eq:auditL3}
\mathcal L_3=H(\phi,X)G^{ab}Y_{ab},\quad
X=-\tfrac12x_ax^a,\quad Y_{ab}=\nabla_a\nabla_b\phi,
\end{equation}
where $H$ denotes $F_{(3)}$ in this appendix. This check keeps $H_X$, arbitrary spacetime curvature and all coefficient derivatives. It does not assume a background geometry, the field equations, or the selection conditions.

\paragraph{1. Separate the three sources of metric variation.}
Let $h^{ab}=\delta g^{ab}$ and hold the covariant fields $x_a,Y_{ab},R_{ab}$ fixed when differentiating explicit inverse-metric factors. Since $\delta X=-x_ax_bh^{ab}/2$, direct differentiation of $H G^{cd}Y_{cd}$ gives
\begin{align}\label{eq:auditfixed}
M^{(3)}_{ab}={}&H\left(2R_{c(a}Y_{b)}{}^c -\tfrac12RY_{ab}-\tfrac12 tR_{ab}\right) -\tfrac12 H_X(G^{cd}Y_{cd})x_ax_b,\\
\mathcal P^{ab}_{(3)}={}&H G^{ab},\quad E^{ab}=H\left(Y^{ab}-\tfrac12t g^{ab}\right).
\end{align}
Here $M^{(3)}_{ab}$ is the fixed-$Y,R_{cd}$ metric derivative, $\mathcal P_{(3)}$ multiplies $\delta Y$, and $E$ multiplies $\delta R_{ab}$. In particular, $H_X$ has not been dropped: it is part of $M^{(3)}_{ab}$, while derivatives of $H$ remain inside the covariant derivatives of $\mathcal P_{(3)}$ and $E$ below.

With $\delta\phi=0$, the connection and Ricci variations are
\begin{align}\label{eq:auditpalatini}
\delta\Gamma^c{}_{ab}&=-\tfrac12\left(\nabla_ah_b{}^c+
\nabla_bh_a{}^c-\nabla^ch_{ab}\right),\\
\delta Y_{ab}&=-x_c\delta\Gamma^c{}_{ab},\quad \delta R_{ab}=\nabla_c\delta\Gamma^c{}_{ab}-\nabla_b\delta\Gamma^c{}_{ac}.
\end{align}
The variation of the {density}, with the metric-volume term included, is therefore
\begin{equation}\label{eq:auditdelta}
\delta(\sqrt{-g}\mathcal L_3)=\sqrt{-g}\left[ \left(M^{(3)}_{ab}-\tfrac12g_{ab}\mathcal L_3\right)h^{ab} +\mathcal P^{ab}_{(3)}\delta Y_{ab}+E^{ab}\delta R_{ab}\right].
\end{equation}
No curvature commutator has been used at this stage.

\paragraph{2. Integrate the connection variation by parts.}
Using Eq.~\eqref{eq:auditpalatini} and the symmetry of $\mathcal P_{(3)}^{ab}$ gives
\begin{equation}\label{eq:auditIY}
\int\!\sqrt{-g}\,\mathcal P^{ab}_{(3)}\delta Y_{ab}
=\int\!\sqrt{-g}\left[-\nabla_c(x_{(a}\mathcal P_{b)}{}^c)
+\tfrac12\nabla^c(x_c\mathcal P_{ab})\right]h^{ab},
\end{equation}
up to a total divergence. For the Ricci term, insertion of the same $\delta\Gamma$ yields, {before} the second integrations by parts,
\begin{align}\label{eq:auditDR}
E^{ab}\delta R_{ab}
={}&-E^{ab}\nabla_c\nabla_a h_b{}^c
+\tfrac12 E^{ab}\Box h_{ab}
+\tfrac12E^{ab}\nabla_a\nabla_b h,\quad h=g_{ab}h^{ab}.
\end{align}
The order $\nabla_c\nabla_a h_b{}^c$ is inherited directly from the Palatini identity. Integrating its first term by parts first in $c$ and then in $a$ gives $-(\nabla_a\nabla_c E^{ab})h_b{}^c$. Here $a$ is summed while $c$ is the free index of the metric variation. Relabelling these indices in the symmetric coefficient of $h^{ab}$ gives $-\nabla_c\nabla_{(a} E_{b)}{}^c$, not the expression with derivatives reversed. The other two terms give $\Box E_{ab}/2$ and $g_{ab}\nabla_c\nabla_d E^{cd}/2$. Thus
\begin{align}\label{eq:auditIR}
I^R_{ab}={}&\tfrac12\Box E_{ab}
-\nabla_c\nabla_{(a} E_{b)}{}^c
+\tfrac12g_{ab}\nabla_c\nabla_dE^{cd}.
\end{align}
This derivation is valid for arbitrary curvature. For clarity, an otherwise similar expression with the derivatives in the middle term reversed is {not} identical: the difference is
\begin{equation}\label{eq:curvaturecommutator}
\left[\tfrac12\Box E_{ab}-\nabla_{(a}\nabla_cE_{b)}{}^c
+\tfrac12g_{ab}\nabla_c\nabla_dE^{cd}\right]-I^R_{ab}
=[\nabla_c,\nabla_{(a}]E_{b)}{}^c.
\end{equation}
With $[\nabla_a,\nabla_b]V^c=R^c{}_{dab}V^d$, this difference is explicitly
\begin{equation}\label{eq:auditcommexpanded}
[\nabla_c,\nabla_{(a}]E_{b)}{}^c
=R_{d(a}E_{b)}{}^d-R^d{}_{(b|c|a)}E_d{}^c.
\end{equation}
It contains independent Ricci-$E$ and Riemann-$E$ contractions and does not generally vanish. Keeping the original Palatini derivative order is essential in a complete unrestricted curvature response.

\paragraph{3. Assemble the complete variation and check it.}
Combining Eqs.~\eqref{eq:auditfixed}, \eqref{eq:auditIY} and \eqref{eq:auditIR} gives the exact bulk metric Euler-Lagrange contribution
\begin{equation}\label{eq:auditmetricEL}
\frac{1}{\sqrt{-g}}\frac{\delta S_3}{\delta g^{ab}}
=M^{(3)}_{ab}+I^{Y(3)}_{ab}+I^R_{ab}
-\tfrac12g_{ab}\mathcal L_3.
\end{equation}
The derivatives of $H(\phi,X)$ occur both algebraically through $H_X$ in Eq.~\eqref{eq:auditfixed} and through the differentiated $H$ in $E^{ab}$ and $\mathcal P_{(3)}^{ab}$. The curvature dependence enters algebraically through $R_{ab},G_{ab}$ and through the commutators implicit in Eq.~\eqref{eq:auditIR}; nothing has been specialized to a flat background.

Two exact identities provide cross-checks. The contracted Bianchi identity gives, for constant $H$,
\begin{equation}\label{eq:auditboundary}
H G^{ab}\nabla_a\nabla_b\phi=\nabla_a(HG^{ab}\nabla_b\phi),
\end{equation}
so the complete metric response must vanish after the volume, algebraic and both connection contributions are added. More generally, varying $\phi$ at fixed $g_{ab}$ gives
\begin{equation}\label{eq:auditscalareq}
\mathcal E^{(3)}_\phi
=H_\phi G^{ab}Y_{ab}
+\nabla_a\left[H_X(G^{cd}Y_{cd})x^a\right]
+\nabla_a\nabla_b(HG^{ab}),
\end{equation}
which enters the diffeomorphism identity
\begin{equation}\label{eq:auditnoether}
2\nabla^a\left(\frac{1}{\sqrt{-g}}\frac{\delta S_3}{\delta g^{ab}}\right)
+\mathcal E^{(3)}_\phi\nabla_b\phi=0.
\end{equation}
These are covariant checks on the complete tensor, not substitutes for its Palatini derivation. The constant-$H$ limit is particularly discriminating: a direct evaluation retaining nonzero Ricci curvature and a nontrivial scalar Hessian cancels in Eq.~\eqref{eq:auditmetricEL}, whereas reversing the middle derivatives of Eq.~\eqref{eq:auditIR} leaves precisely the curvature commutator in Eq.~\eqref{eq:curvaturecommutator}. Independently, the fixed-$Y,R_{cd}$ derivative in Eq.~\eqref{eq:auditfixed} can be verified by differentiating $H[g^{ac}g^{bd}R_{cd}Y_{ab}-(g^{ab}R_{ab})(g^{cd}Y_{cd})/2]$ with respect to a symmetric inverse-metric variation at fixed covariant tensors. Equations~\eqref{eq:auditfixed}-\eqref{eq:auditIR} specify the complete off-shell covariant variation, including nonzero curvature and the derivatives of $H(\phi,X)$.

\section{Complete metric response and its fluid projections}\label{app:fullmetric}
To compute the remaining contributions, introduce the six contractions defined in Sec.~\ref{sec:DHOSTbasis} and $v_a=Y_{ab}x^b$, $w_a=Y_{ab}v^b$, $\ell_a=Y_{ab}w^b$. Put $j_t=\partial\mathcal J_0/\partial t$, and similarly for $U,V,S,W,Z_3$. Their full Wilson-function dependence is displayed in Eq.~\eqref{eq:js}; the following expressions determine both the metric and connection contributions without a further variation:
\begin{subequations}\label{eq:fullM}
\begin{align}
M^{(0)}_{ab}={}&-\frac12\mathcal J_{0,X}x_ax_b+j_tY_{ab}
+2j_U(Y^2)_{ab}+3j_V(Y^3)_{ab}+2j_Sx_{(a}v_{b)}\\
&+j_W\big(v_av_b+2x_{(a}w_{b)}\big)
+j_{Z_3}\big(2x_{(a}\ell_{b)}+2v_{(a}w_{b)}\big),\label{eq:M0full}\\
M^{(3)}_{ab}={}&F_{(3)}\big[2R_{c(a}Y_{b)}{}^c-\tfrac12RY_{ab}-\tfrac12tR_{ab}\big]
-\tfrac12F_{(3),X}(G^{cd}Y_{cd})x_ax_b,\label{eq:M3full}\\
\Mom^{ab}={}&j_tg^{ab}+2j_UY^{ab}+3j_V(Y^2)^{ab}+j_Sx^ax^b
+2j_Wx^{(a}v^{b)}+j_{Z_3}\big(2x^{(a}w^{b)}+v^av^b\big)
+F_{(3)}G^{ab}.\label{eq:momentumall}
\end{align}
\end{subequations}
In $\mathcal J_{0,X}$ the invariants $t,U,V,S,W,Z_3$ are held fixed; hence it differentiates the Wilson coefficients only. The derivatives of the explicit metrics have already been evaluated in the other terms. $M_{ab}=M^{(0)}_{ab}+M^{(3)}_{ab}$, and inserting the displayed $\Mom^{ab}$ into Eq.~\eqref{eq:IY} produces the entire Hessian-connection response. For constant $Q$, the full identity is
$M_{ab}^{(Q)}+I_{ab}^{Y(Q)}=\frac12(Qt)g_{ab}$, not zero separately. Thus the variation
$\frac12(Qt)g_{ab}-M_{ab}^{(Q)}-I_{ab}^{Y(Q)}$ vanishes exactly, as required for a boundary interaction. This explicitly checks the volume-form contribution as well as the connection variation.

For an explicit $1+3$ evaluation, decompose this already-calculated momentum as
\begin{equation}\label{eq:momentumframe}
\Mom_{ab}=m u_au_b-2u_{(a}m_{b)}+m_{ab},\quad
m=\Mom_{uu},\quad m_a=h_a{}^cu^d\Mom_{cd},\quad
m_{ab}=h_a{}^ch_b{}^d\Mom_{cd},\quad \tau=h^{ab}m_{ab}.
\end{equation}
All $m$'s are fixed algebraic polynomials in the Wilson functions and $Y_{ab}$, through Eq.~\eqref{eq:momentumall}; they are not thermodynamic fitting parameters. The derivatives in the connection term then have the exact projections
\begin{subequations}\label{eq:Iallproj}
\begin{align}
u^au^bI^Y_{ab}={}&-\frac12\nabla_c(smu^c)+\nabla_c(sm^c)-sK^{ab}m_{ab},\label{eq:Iuu}\\
h^{ab}I^Y_{ab}={}&\frac12\nabla_c(s\tau u^c)-sK^{ab}m_{ab},\label{eq:Ihh}\\
I^Y_{\langle ab\rangle}={}&\frac12(d+s\theta)m_{\langle ab\rangle}
+\frac{s}{2}\big(h_a{}^ch_b{}^d\dot m_{cd}\big)_{\!\mathrm{TF}}
-s\big(K_{(a}{}^cm_{b)c}\big)_{\!\mathrm{TF}}.\label{eq:Itf}
\end{align}
\end{subequations}
The mixed component $h_a{}^bu^cI^Y_{bc}$ was given in Eq.~\eqref{eq:connection}. All these expressions follow by differentiating Eq.~\eqref{eq:momentumframe} and using $\nabla_a u_b=K_{ab}-u_a a_b$. The three displayed lines supply the projections needed for energy density, pressure and anisotropic stress; the mixed projection is Eq.~\eqref{eq:connection}.

\section{Derivation of the algebraic invariant responses}\label{app:derivationInvariants}
For a compact verification of Eq.~(\ref{eq:Aexplicit}), temporarily hold $\phi,X$ and every Wilson function fixed. Symmetric variation of the inverse metric at fixed covariant $x_a,Y_{ab}$ gives
\begin{align}
\frac{\partial t}{\partial g^{ab}}&=Y_{ab},&\frac{\partial U}{\partial g^{ab}}&=2(Y^2)_{ab},\notag\\
\frac{\partial V}{\partial g^{ab}}&=3(Y^3)_{ab},&\frac{\partial S}{\partial g^{ab}}&=2x_{(a}v_{b)},\notag\\
\frac{\partial W}{\partial g^{ab}}&=v_av_b+2x_{(a}w_{b)},\notag\\
\frac{\partial Z_3}{\partial g^{ab}}&=2x_{(a}\ell_{b)}+2v_{(a}w_{b)}.
\end{align}
Multiplying each line by its $j$ in Eq.~(\ref{eq:js}) yields the fixed-Wilson part of $M^{(0)}_{ab}$. Every additional term proportional to $\partial_X\mathcal J_0$ is $-\tfrac12(\partial_X\mathcal J_0)x_ax_b$, so its mixed projection vanishes. Replacing the Hessian with Eq.~(\ref{eq:hesskin}) and using $h_a{}^bu^c u_{(b}V_{c)}=-V_a/2$ for spatial $V_a$ reproduces Eq.~(\ref{eq:Aexplicit}) term by term.

\section{Effective-fluid projections of the curvature, kinetic and braiding sectors}\label{app:FPQprojection}
\begin{subequations}\label{eq:allFPQ}
\begin{align}
F_{(2)}\rho={}&\tfrac12(2XP_X-P)+XQ_\phi+sX\theta Q_X+\D^2F_{(2)}-\theta\dot F_{(2)}+XRF_{(2),X},\\
F_{(2)}p={}&\tfrac12P+XQ_\phi-XQ_Xd+\ddot F_{(2)}-a^a\D_aF_{(2)}-\tfrac23\D^2F_{(2)}+\tfrac23\theta\dot F_{(2)},\\
F_{(2)}q_a={}&-\D_a\dot F_{(2)}+K_a{}^b\D_bF_{(2)}-sXQ_Xa_a,\\
F_{(2)}\pi_{ab}={}&\D_{\langle a}\D_{b\rangle}F_{(2)}-\dot F_{(2)}\sigma_{ab}.
\end{align}
\end{subequations}

After the full variation, the $P+Qt$ sector can be collected in the manifestly symmetric tensor
\begin{equation}
\begin{aligned}\mathcal S^{P+Q}_{ab}={}&\tfrac12P g_{ab}+\tfrac12P_Xx_ax_b\\
&+\tfrac12Q_Xt x_ax_b+x_{(a}\nabla_{b)}Q\\
&-\tfrac12g_{ab}x^c\nabla_cQ.\end{aligned}
\end{equation}
Here the full $Qt g_{ab}/2$ and $QY_{ab}$ terms canceled against the connection variation. From $\nabla_aX=-Y_{ab}x^b$ and Eq.~(\ref{eq:hesskin}),
\begin{align}
\nabla_aX&=-sd\,u_a-s^2a_a,\notag\\
 x^a\nabla_aQ&=-s^2Q_\phi+s^2dQ_X.
\end{align}
Projecting $\mathcal S^{P+Q}_{ab}$ yields $\rho_{P+Q}$, $p_{P+Q}$ and $q^{P+Q}$ in Eq.~(\ref{eq:allFPQ}), with no trace-free spatial part. For a general scalar $F_{(2)}$ one independently has
\begin{align}
 u^au^b(\nabla_a\nabla_bF_{(2)}-g_{ab}\Box F_{(2)})&=\D^2F_{(2)}-\theta\dot F_{(2)},\\
 \tfrac13 h^{ab}(\nabla_a\nabla_bF_{(2)}-g_{ab}\Box F_{(2)})&=\ddot F_{(2)}-a^c\D_cF_{(2)}\notag\\
 &\quad-\tfrac23\D^2F_{(2)}+\tfrac23\theta\dot F_{(2)},\\
 -h_a{}^bu^c\nabla_b\nabla_cF_{(2)}&=-\D_a\dot F_{(2)}+K_a{}^b\D_bF_{(2)},\\
 (\nabla_a\nabla_bF_{(2)})_{\langle ab\rangle}&=\D_{\langle a}\D_{b\rangle}F_{(2)}-\dot F_{(2)}\sigma_{ab}.
\end{align}
The $RF_{(2),X}x_ax_b/2$ term contributes only $XRF_{(2),X}$ to $F_{(2)}\rho$. Adding these pieces proves all four equations (\ref{eq:allFPQ}).

\section{Kinetic degeneracy of the Eckart-selected interactions}\label{app:kineticdegeneracy}

In this appendix, we derive the kinetic determinant~\eqref{eq:kineticDeterminantFactored} used in Sec.~\ref{sec:covariantHeatConnection}, following the arbitrary-foliation approach of the DHOST literature~\cite{Langlois:2015cwa,BenAchour:2016cay,BenAchour:2016fzp,Langlois:2018dxi}. The calculation makes explicit how the rank condition obtained independently from Eckart matching restricts the kinetic structure of the selected interactions.

Consider the reduced Lagrangian $F(\phi)R+P(\phi,X)+H(\phi,X,t,S)$ and the scalar-gradient decomposition~\eqref{eq:tiltedADMnormal}. From Eq.~\eqref{eq:boostedVelocities}, the velocity dependence of the two Hessian contractions can be written as
\begin{equation}
\begin{pmatrix}
t\\ S
\end{pmatrix}_{\!\rm vel}
=\ell_v v+\ell_K\mathbf K,
\quad
\ell_v=\begin{pmatrix}-1\\ A_*^2\end{pmatrix},
\end{equation}
where $\mathbf K$ denotes the six independent components of $K^{(n)}_{ij}$ and $\ell_K$ is the corresponding $2\times6$ matrix of metric-velocity coefficients.

Defining the Hessian matrix
\begin{equation}
C=\begin{pmatrix}
H_{tt}&H_{tS}\\
H_{tS}&H_{SS}
\end{pmatrix},
\end{equation}
with derivatives evaluated at fixed $(\phi,X)$, the coupled scalar-metric kinetic matrix takes the form
\begin{equation}\label{eq:explicitKineticBlock}
\mathbb K=
\begin{pmatrix}
\ell_v^TC\ell_v & \ell_v^TC\ell_K\\
\ell_K^TC\ell_v & \mathbb G+\ell_K^TC\ell_K
\end{pmatrix},
\end{equation}
where $\mathbb G$ is the gravitational DeWitt kinetic block associated with $F(\phi)R$, which is nonsingular for $F\ne0$. The remaining terms arising from $P$, $Qt$, and derivatives of $F(\phi)$ do not contribute to the kinetic Hessian in these highest-velocity variables.

The thermodynamic integrability condition~\eqref{eq:generalRankCondition} requires $\det C=0$. For a nonvanishing rank-one Hessian, we can therefore write $C=\rho rr^T$, where $\rho$ is a scalar and $r$ is a two-component vector. Introducing $\alpha=\ell_v^Tr$ and $\beta=\ell_K^Tr$, Eq.~\eqref{eq:explicitKineticBlock} becomes
\begin{equation}
\mathbb K=
\begin{pmatrix}
\rho\alpha^2 & \rho\alpha\beta^T\\
\rho\alpha\beta & \mathbb G+\rho\beta\beta^T
\end{pmatrix}.
\end{equation}
When the lower-right block is invertible, the Schur complement and the matrix determinant lemma give
\begin{align}
\det\mathbb K
&=\det(\mathbb G+\rho\beta\beta^T)
\frac{\rho\alpha^2}{1+\rho\beta^T\mathbb G^{-1}\beta},\\
\det(\mathbb G+\rho\beta\beta^T)
&=\det\mathbb G\left(1+\rho\beta^T\mathbb G^{-1}\beta\right).
\end{align}
Combining these expressions yields
\begin{equation}
\det\mathbb K=\det\mathbb G\,\rho\alpha^2
=\det\mathbb G\,\ell_v^TC\ell_v.
\end{equation}
Since both sides are polynomial expressions in the kinetic matrix components, the identity extends by continuity to configurations where the lower-right block becomes singular. It also holds when $C=0$.

Substituting the explicit expression for $\ell_v$ gives
\begin{equation}
\det\mathbb K=\det\mathbb G\left(H_{tt}-2A_*^2H_{tS}+A_*^4H_{SS}\right),
\end{equation}
which establishes Eq.~\eqref{eq:kineticDeterminantFactored}. The importance of the thermodynamic rank condition is that it removes the additional determinant contribution that would otherwise arise from a rank-two Hessian response.

At fixed $(\phi,X,t,S)$, covariant degeneracy requires the remaining kinetic factor to vanish for an open set of scalar tilts, with $A_*^2=2X+p^2$. Since this factor is polynomial in $A_*^2$, its coefficients must vanish independently, giving $H_{tt}=H_{tS}=H_{SS}=0$, in agreement with Eq.~\eqref{eq:fullCovariantHealth}.

Thus, although the general DHOST framework admits nonlinear scalar-Hessian interactions through cancellations in the coupled kinetic matrix, the independent restrictions imposed by Eckart matching eliminate the possibility of such cancellations within the selected family. The surviving covariantly degenerate interactions consequently belong to the established Horndeski subclass~\eqref{eq:generalHealthyIntersection}.

\end{document}